\documentclass[showpacs,preprintnumbers,amsmath,amssymb,latexsym,array,enumerate,letter]{revtex4}
\usepackage{graphicx}
\usepackage{dcolumn}
\usepackage{bm}
\usepackage{epsfig}
\usepackage{slashed}

\newcommand{\half}{\frac{1}{2}}
\newcommand\beq{\begin{eqnarray}}
\newcommand\eeq{\end{eqnarray}} 
\newcommand\eqn[1]{\label{eq:#1}} 
\newcommand\eq[1]{Eq.~(\ref{eq:#1})} 
\newcommand{\vev}[1]{\langle #1 \rangle}

\newcommand{\GeV}{{\rm ~GeV }}
\newcommand{\TeV}{{\rm ~TeV }}
\newcommand{\MeV}{{\rm ~MeV }}

\newcommand{\CH}{{\cal H}}
\newcommand{\CO}{{\cal O}}
\newcommand{\CP}{{\cal P}}
\newcommand{\CM}{{\cal M}}
\newcommand{\CN}{{\cal N}}

\newcommand{\CR}{{\cal R}}

\newcommand{\CY}{{\cal Y}}

\newcommand{\CL}{{\cal L}}

\newcommand{\Tr}{{\rm Tr\,}}

\newcommand{\mybar}[1]%
        {\kern  .6pt\overline{\kern - .6pt#1\kern - .6pt}\kern  .6pt}
\begin{document}

\preprint{INT-PUB-13-010, NSF-KITP-13-037}

\title{Little flavor: A model of weak-scale flavor physics}

\author{Sichun Sun}
 \email{sichun@uw.edu}
\affiliation{Institute for Nuclear Theory, Box 351550, Seattle, WA 98195-1550, USA}

 \affiliation{Kavli Institute for Theoretical Physics, University of California Santa Barbara, CA 93106-4030, USA}

\author{David B. Kaplan}
\email{dbkaplan@uw.edu}
\affiliation{Institute for Nuclear Theory, Box 351550, Seattle, WA 98195-1550, USA}
 
\author{Ann E. Nelson}
\email{anelson@uw.edu}

 \affiliation{Dept. of Physics, University of Washington, Box 351560, Seattle, WA 98195-1560, USA}
 
 \date{\today}
 
 \begin{abstract}
 We describe a model of quarks which identifies the large global symmetries of little Higgs models with the global flavor symmetries that arise in a deconstruction of the extra-dimensional ``topological insulator" model of flavor. The nonlinearly realized symmetries  of little Higgs theories play a critical role in determining the  flavor  structure of fermion masses and mixing.   All of flavor physics occurs at the few  TeV scale in this model, yet flavor changing neutral currents arising from the new physics are naturally smaller than those generated radiatively in the standard model, without having to invoke minimal flavor violation.  
 
  \end{abstract}

\maketitle

\section{Introduction}
The standard model (SM) is extremely successful at predicting what we do not see --- namely flavor changing neutral currents (FCNC), lepton family violation among charged leptons, proton decay or neutron oscillations, and (with the exception of the strong $CP$ problem)  large $CP$ violating effects.  These all follow from the fact that such processes require irrelevant operators in the SM and are therefore suppressed by the high energy scale associated with new heavy particles.  By assuming a desert for many decades of energy above the electroweak scale, all of the above processes are strongly suppressed, providing a simple explanation for what we (don't) see.  The SM is unsatisfying at the same time, as the hierarchical structure of fermion families is put in by hand with no explanation.  An interesting generic explanation for flavor structure was posited long ago by Froggatt and Nielsen \cite{Froggatt:1978nt}, in which large approximate flavor symmetries are broken hierarchically by multiple spurions,  which individually break the flavor symmetry, but none by itself sufficiently breaking the symmetry to provide Yukawa couplings for all the SM fermions.  Since then, many models of flavor have been built on this premise; however, with a desert above the electroweak scale to explain the absence of FCNC and electric dipole moments, it would appear that experimental clues to the origins of fermion family structure would be well beyond the reach of any foreseeable experiment, and so this scientific program has remained inconclusive and unconvincing.  

There is tension in the SM, however, between the natural explanation of a desert for the absence of FCNC, lepton and baryon number violation, and $CP$ violation on the one hand, and the fine tuning of the Higgs sector that comes with a desert on the other.  There have been numerous attempts to modify the SM to remove this tension.  Walking technicolor, for example, maintains the desert while replacing the Higgs sector of the SM with dynamical symmetry breaking --- but is no longer viable with the discovery of the Higgs.  Another approach, such as in Effective Supersymmetry  \cite{Cohen:1996vb}, is to  populate the desert while maintaining enough approximate symmetries that suppress the dangerous FCNC and symmetry violating processes.  These theories all attempt to extend the viability of the SM up to the GUT scale.  However an interesting and relatively recent alternative is the Little Higgs Mechanism, which  extends naturalness in the SM only up to the $\sim 10\TeV$ scale \cite{ArkaniHamed:2001nc,ArkaniHamed:2002qy,ArkaniHamed:2002qx,Schmaltz:2005ky}.  In these models, composite Higgs theories \cite{Kaplan:1983fs,Kaplan:1983sm,Georgi:1984ef,Georgi:1984af,Dugan:1984hq} are designed with large nonlinearly realized symmetries broken by sparse spurions, none of which by themselves break the symmetries sufficiently to allow a Higgs potential to be radiatively generated at one loop\footnote{A previous model  which addressed the origin of fermion masses at the TeV scale  without invoking minimal flavor violation,   ref. \cite{Nelson:1988wn}, invoked a very large global symmetry group and sparse set of spurions to achieve a flavor hierarchy. This model demonstrated that the same symmetries that produced a fermion mass hierarchy could suppress FCNC.   In ref. \cite{ ArkaniHamed:1999yy}, the extra-dimensional  mechanism of ``shining" was introduced to produce the fermion masses at the TeV scale without FCNC. Extra-dimensional theories, deconstructed \cite{ ArkaniHamed:2001ca} automatically provide the very large approximate symmetry group and sparse set of spurions which can be useful for flavor models, as well as for little Higgs models.}.

It is intriguing that the underlying mechanism of the Little Higgs Mechanism is similar in spirit to the Froggatt-Nielsen program for flavor structure, even if applied in a  different way to a different problem.  In this paper we present an effective theory valid up to  the $\sim  20\TeV$ scale model that exhibits a large approximate global symmetry broken by means of sparse spurions which combine to   give rise to both electroweak symmetry breaking, as well as the observed hierarchies of quark masses and mixing angles.  Although we hope this approach may lead to a deep understanding of flavor, the model  we present is less ambitious, reproducing the SM quark masses and mixings without predicting them, and not addressing the leptonic sector of the SM. The point of the model  is  to demonstrate that flavor physics can lie just beyond the electroweak symmetry breaking scale --- and can be intimately related to it --- without giving rise  to FCNC or electric dipole moments   in conflict with experiment, and without assuming Minimal Flavor Violation \cite{Chivukula:1987py}.  We show that such a theory, fit to give the observed quark masses and CKM angles to within a few percent, has rich phenomenology with exotic quarks, mesons and massive gauge bosons at the few $\TeV$ scale.

We begin by explaining the general structure and symmetries of the model, which consist of an approximate $U(3)$ flavor symmetry times a product of approximate $SU(4)$ symmetries in which is embedded the $SU(2)\times U(1)$ gauge group of the SM.  Some of the $SU(4)$ symmetries are nonlinearly realized, and two Higgs doublets appear as pseudo Nambu-Goldstone bosons of an $SU(4) \times SU(4)/SU(4)$  nonlinear sigma model. Although the $SU(4)$ groups  are not family symmetries, explicit $SU(4)$ breaking by spurions is required before the SM quarks can obtain nonzero Yukawa couplings to the Higgs, with a nontrivial structure arising from the simultaneous breaking of the $U(3)$ family symmetry.   After discussing the structure of quark masses and mixing angles,  we provide an explicit fit to existing data and show how FCNC in this fit are well within experimental limits.   Next we turn to the Higgs potential; by construction the Little Higgs mechanism is at work in eliminating divergent radiative corrections from one-loop fermion contributions.  We then briefly explain how our model is inspired by the deconstruction of the extra dimension domain wall fermion/topological insulator model of flavor of reference \cite{Kaplan:2011vz}, and conclude with a discussion of  how our approach might be extended.

\section{The $SU(4)\times U(3)$ Little Flavor Model}

Our model  is characterized by the moose diagram in Fig.~\ref{fig:sixsite}, consisting of six sites, three white and three black, connected by oriented links.  Fermions live on the sites and mesons on the links, while some gauge bosons reside only on the white sites and others only on the black sites.

\subsection{Gauge symmetries and the Higgs}

The gauge symmetry of the model is $SU(3)\times G_w\times G_b$, where $SU(3)$ is color and  $G_{w,b}$ are independent $SU(2)\times U(1)$ groups associated with white (w) and black (b) sites respectively. The SM electroweak gauge group is the diagonal subgroup of $G_w\times G_b$, and  we take the the gauge couplings to be
\beq
g_{1,w}=\frac{g'}{\cos\gamma_1}\ ,\qquad g_{1,b} = \frac{g'}{\sin\gamma_1}\ ,\qquad 
g_{2,w}=\frac{g}{\cos\gamma_2}\ ,\qquad g_{2,b} = \frac{g}{\sin\gamma_2}\ .
\eqn{gamma}\eeq
 where $g=e/\sin\theta_w$ and $g'=e/\cos\theta_w$ are the usual SM gauge couplings and the angles $\gamma_{1,2}$ are free parameters.  

The gauge fields are coupled to  an $SU(4)\times SU(4)/SU(4)$ nonlinear $\sigma$-model, parametrized by the field $\Sigma$, an $SU(4)$ matrix which transforms under $SU(4)\times SU(4)$ as the $(4,\mybar 4)$ representation.  The $G_w\times G_b$ gauge symmetry is embedded in the $SU(4)\times SU(4)$ so that the covariant derivative acts on $\Sigma$ as
\beq
D_\mu\Sigma = \partial_\mu\Sigma +i\left(g_{2,w}A^a_\mu T_a + g_{1,w}  B_\mu Y\right) \Sigma - i\Sigma \left(g_{2,b}\tilde A^a_\mu T_a+g_{1,b} \tilde B_\mu Y\right)\ ,
\eeq
where $\{A^a_\mu, B_\mu\}$ and $\{\tilde A^a_\mu, \tilde B_\mu\}$ are the gauge bosons of $G_w$ and $G_b$ respectively, while the generators can be written in a $2\times 2$ block notation as
\beq
T_a = \half \begin{pmatrix} \sigma_a & 0\\0 & 0\end{pmatrix}\ ,\qquad
Y =  \begin{pmatrix} 0 & 0\\0 & T_3\end{pmatrix} \ .
\eqn{tdef}\eeq
The $\Sigma$ field breaks $G_w\times G_b$ gauge symmetry down to a diagonal subgroup; if $\vev{\Sigma}=1$, the unbroken subgroup is the diagonal $SU(2)\times U(1)$, which is identified with the electroweak gauge group of the SM, and it has the correct couplings $g$ and $g'$. The spectrum then contains two exotic $Z$ bosons and an exotic $W$ boson, whose masses are given by
\beq
M_{Z'} = M_{W'}=\frac{g f}{\sin 2\gamma_2}\ ,\qquad
M_{Z''} = \frac{g' f}{\sin 2\gamma_1}\ .
\eqn{heavyZ}\eeq
Electroweak symmetry breaking will correct these relations at $O(M^2_Z/f^2)$; in the model we consider in this paper we fix the Goldstone boson decay constant to be $f=1.5\TeV$; thus the corrections are $O(M^2_Z/f^2)\simeq 1\%$.

The $\Sigma$ field describes fifteen pseudo Goldstone bosons with  decay constant $f$, to be set to $1.5\TeV$ in the phenomenological model we describe below.  It can be conveniently parametrized as
\beq
\Sigma = \xi \,\xi_\eta\,\xi_\pi\,\Sigma_H\,\xi_\pi\,\xi_\eta\,\xi\ ,
\eeq
where
\beq
\xi &=& \exp \left[\frac{i}{2f}\begin{pmatrix} \vec\pi\,'\cdot\vec\sigma & 0\cr 0 & \pi_3 \sigma_3 \end{pmatrix}  \right] 
\eqn{eaten}
\eeq
\beq
\xi_\eta =  \exp \left[\frac{i}{\sqrt{8}f}\begin{pmatrix}  \eta  &\cr & -\eta  \end{pmatrix}  \right]  \, ,
\eqn{eta}
\eeq
\beq
\xi_\pi = \exp \left[\frac{i}{\sqrt{2}f}\begin{pmatrix}   0& \cr  &\Pi \end{pmatrix}  \right]  \, , \quad \Pi \equiv \begin{pmatrix}0&\pi^+\cr \pi^- &0\end{pmatrix}\ .
\eqn{pi}
\eeq
\beq
\Sigma_H=\exp\left[(i\sqrt{2}/f)\begin{pmatrix}0 & -i\CH^\dagger \cr i\CH& 0 \end{pmatrix}\right] \ ,\eqn{Higgs}
\eeq
The field $\xi$ contains the Goldstone bosons eaten when $G_w\times G_w$ is broken to the diagonal $SU(2)\times U(1)$, and in unitary gauge it is rotated away.  The $\eta$ and $\pi^\pm$  fields correspond to exotic $SU(2)$ singlets which are neutral and charged respectively.  Finally,    $\CH$ contains two electroweak doublets which will be identified with the two SM Higgs doublets, $H_u$ and $H_d$:
 \beq
\CH =\begin{pmatrix}- H_u^T\epsilon\cr \ \ H_d^T\epsilon\end{pmatrix}  =  \begin{pmatrix} h_u^0 & -h_u^+\cr -h_d^-& h_d^0\end{pmatrix} \ .
\eeq
The potential for $\Sigma$ will cause a small misalignment away from the $SU(2)\times U(1)$ preserving vacuum $\vev{\Sigma}=1$, corresponding to nonzero vevs of the Higgs doublets, an example of the composite Higgs mechanism \cite{Kaplan:1983fs,Kaplan:1983sm,Georgi:1984ef,Georgi:1984af,Dugan:1984hq} (see references \cite{Agashe:2005dk,Panico:2012uw,Barbieri:2012tu}  for some more recent developments in composite Higgs theories).  Assuming $\vev{h_u^0}=v_u/\sqrt{2}$ and $\vev{h_d^0}=v_d/\sqrt{2}$, the electroweak breaking vacuum corresponds to
\beq
\vev{\Sigma} = \begin{pmatrix} 
c_u &  0 & s_u & 0\cr
  0 &   c_d & 0 & s_d\cr
-s_u &  0 & c_u & 0\cr
  0 & -s_d & 0 & c_d
\end{pmatrix}\ ,
\qquad
c_{u,d} = \cos \frac{v_{u,d}}{f}\ ,\quad
s_{u,d} = \sin \frac{v_{u,d}}{f}\ .
\eqn{ewsb}\eeq
In the special case $v_u=v_d=v$,  (or $\tan\beta=1$) obtaining the correct $W$ and $Z$ masses requires
\beq
\sin\frac{v}{f} = \frac{ M_Z\sin2\theta_w }{\sqrt{2} e f}\ ,
\eeq
with additional corrections of size $O(M^2_Z/f^2)\simeq 1\%$.

The interactions of the mesons are described by a chiral Lagrangian defined with a momentum cutoff at the scale $\Lambda \sim 4\pi f\sim 19 \TeV$ in the model we describe here. The leading operator is given by
\beq
\frac{f^2}{4} \Tr(D_\mu\Sigma)^\dagger D^\mu\Sigma\ ,
\eeq
which gives canonically normalized meson fields for $\vev{\Sigma}=1$, but care must be taken to account for $O(v/f)$ corrections to the wavefunction normalization  in the electroweak symmetry breaking vacuum \eq{ewsb}.    

Before describing the potential for $\Sigma$ and its vacuum alignment, we must first discuss the fermions in the model and their Yukawa couplings to $\Sigma$.

\subsection{Fermions}

\begin{figure}[t]
\includegraphics[width=4 cm]{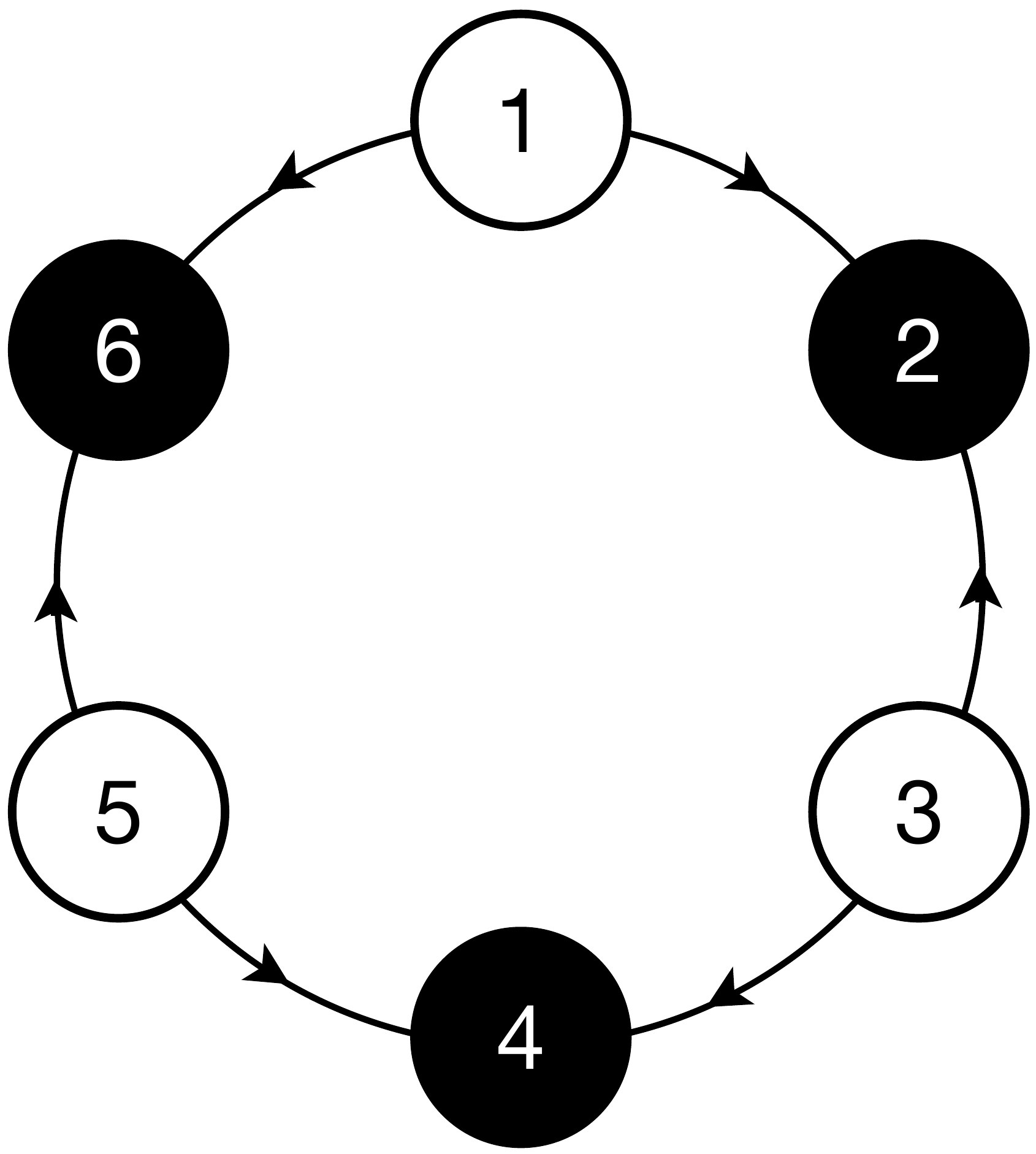}
\caption{ {The moose describing the $SU(4)$ model.  Chiral fermions reside at white sites, and vector fermions at  black sites.The links correspond a nonlinear $\Sigma$ field which contains $H_u$ and $H_d$ Higgs fields, although in the model we construct, $\Sigma$ fields only live on three of the six possible links pictured here. The gauge symmetry of the theory is $SU(3)\times [SU(2)\times U(1)]^2$, where $SU(3)$ is color; the two $SU(2)\times U(1)$ groups are associated with the white sites and the black sites, and the conventional electroweak gauge group  resides in their diagonal subgroup.}}
\label{fig:sixsite}
\end{figure}

Our model is described by the moose diagram of Fig.~\ref{fig:sixsite}, where the oriented links represent $\Sigma$ and $\Sigma^\dagger$, while the white and black sites represent fermions transforming nontrivially under $G_w$ and $G_b$ respectively.
The fermions are all color triplet (we only consider quarks in this model, not leptons)  and consist of
$SU(2)$ doublets $Q = (u,d)$ and $SU(2)$ singlets $(U,\,D)$. The fermions on the black sites are Dirac fermions and are grouped together as the quartets of  approximate $U(4)_b$ symmetries
\beq
\psi_{b} = \begin{pmatrix} Q\\ U\\ D \end{pmatrix}_b\ ,\qquad \text{black sites: }b=2,4,6\ .
\eqn{psi}\eeq
The fermions on the white sites are chiral fermions, conveniently packaged as incomplete quartets of independent approximate  $U(4)_w$ symmetries:
\beq
\chi_{w,L}  =  \begin{pmatrix} Q\\0\\ 0 \end{pmatrix}_{w,L}\ ,\qquad 
\chi_{w,R}=  \begin{pmatrix} 0\\U\\ D \end{pmatrix}_{w,R}\ ,\qquad \text{white sites: }w=1,3,5\ .
\eqn{chi}\eeq
In the above expressions, the subscripts $b=2,4,6$ and $w=1,3,5$ refer to the site numbers in Fig.~\ref{fig:sixsite}.

The $G_w$ and $G_b$ gauge generators  are embedded within $U(4)_{w,b}$ exactly as in \eq{tdef}, except that $Y$  is extended to include a term   $\half(B-L)=\frac{1}{6}$ for all the fermions,  vanishing for the mesons,
\beq
  Y =  \begin{pmatrix} 0 & 0\\0 & T_3\end{pmatrix} + \frac{1}{2} (B-L)=  \begin{pmatrix} 0 & 0\\0 & T_3\end{pmatrix} + \frac{1}{6} \begin{pmatrix} 1 & 0\\0 & 1\end{pmatrix}\qquad\text{(fermions)}
\eqn{embed}
\eeq
where the colored fermions all carry $(B-L)=\frac{1}{3}$. Thus the covariant derivatives act on the fermions as
\beq
D_\mu \chi_{L,w} &=& (\partial_\mu + i g_{2,w}  A^a_{\mu} T_a + ig_{1,w}  B_\mu  Y )\chi_{L,w} \ ,\cr &&\cr
D_\mu \chi_{R,w} &=& (\partial_\mu +  i g_{1,w}  B_\mu Y )\chi_{R,w} \ ,\cr&&\cr
D_\mu \psi_b &=& (\partial_\mu + i g_{2,b} \tilde A^a_\mu T_a + ig_{1,b} \tilde B_\mu Y)\psi_b\ .
\eeq

\subsection{Yukawa couplings and masses}
\label{sec:yukawa}

The masses and Yukawa terms in our model  come in two types:  those that preserve an  $SU(4)\times U(3)$ symmetry, and those where that symmetry is partially broken by spurions.  The $SU(4)$ is the symmetry of the degenerate massive vector fermions on the black sites, which is identified with the $SU(4)_R$ symmetry of the nonlinear $\sigma$-model; the $U(3)$ will be identified as a family symmetry and contains the $S_3$ permutation symmetry of the moose of Fig.~\ref{fig:sixsite}.

\subsubsection{The $SU(4)\times U(3)$ symmetric terms}

To make the $U(3)$ symmetry manifest it is useful to consider the ``unit cell" of our moose digram Fig.~\ref{fig:sixsite} to consist of an adjacent pair of black and white sites, the moose consisting of three such pairs.  We label the cells by $n=1,2,3$, with cell $n$ associated with sites $\{2n-1,2n\}$, and then an index $\alpha=1,2$ will specify the white and the black site respectively within the cell.  The fermions are all labeled then as $\Psi_{n,\alpha}$ with 
\beq
\Psi_{1,1} = \chi_{1} \ ,\quad
\Psi_{1,2} = \psi_{2} \ ,\quad
\Psi_{2,1} = \chi_{3} \ ,\quad
\Psi_{2,2} = \psi_{4} \ ,\quad
\Psi_{3,1} = \chi_{5} \ ,\quad
\Psi_{3,2} = \psi_{6} \ ,
\eeq
where the $\chi$ are the four-component  chiral fermions on the white sites in \eq{chi}, and the $\psi$ are the four component Dirac fermions on the black sites in \eq{psi}.

The symmetric fermion mass and Yukawa terms are given by
\beq
\CL_\text{sym} =  \mybar \Psi_{m\alpha,L} \left[\CM_{m\alpha,n\beta} +\Sigma\, \CY_{m\alpha,n\beta} -\Sigma^\dagger\,\CY^\dagger_{m\alpha,n\beta}\right]  \Psi_{n\beta,R}+ h.c.\ ,
\eqn{Lsym}\eeq
where $\CM^{(0)}$, $\CY$ and $\mybar \CY$ are independent and take the form
\beq
\CM_{m\alpha,n\beta}
= M  \,\begin{pmatrix} 1&&\cr &1&\cr&&1\end{pmatrix}_{mn}\otimes\ \ \begin{pmatrix}0 &0\\ 0& 1\end{pmatrix}_{\alpha\beta}
\eqn{cmval}\eeq
\beq
\CY =
 \lambda\, f\, \begin{pmatrix} 1&&\cr &1&\cr&&1\end{pmatrix}_{mn}\otimes\ \ \begin{pmatrix}0 &1\\ 0& 0\end{pmatrix}_{\alpha\beta} \ ,
\eqn{cyval}\eeq
where all unmarked matrix elements are zero. We have written the mass and Yukawa couplings in a direct product notation to make manifest the $U(3)$ symmetry acting on the unit cell indices $m,n$.
The $\CM$ term is a common mass term for the black site Dirac fermions; the $\CY$ term is a nearest neighbor hopping interaction involving $\Sigma$ in the direction of the link arrow, from white site to black site within the cell, and the $\CY^\dagger$ term is a hopping interaction against the link arrow, from black to white, involving $\Sigma^\dagger$; combined these hopping terms look like a  covariant derivative in a fifth dimension, with $\Sigma$ playing the role of the fifth component of a gauge field.  Having the hopping strength be the same in the forward and backward directions is protected by a discrete $Z_2$ symmetry. Note though that the $\Sigma$ field only acts on the three links that connect black and white sites within a cell; in this model we do not have $\Sigma$ fields acting on the links between cells.

Less obvious in this notation is that $\CL_\text{sym}$ is invariant under a nonlinearly realized  $SU(4)_L\times SU(4)_R'$ symmetry which is the $SU(4)_L$ symmetry of the $\sigma$-model, times the diagonal subgroup of the $\sigma$-model's $SU(4)_R$ and the vector $SU(4)$ symmetry of the black site Dirac fermions.  A remarkable consequence of this $SU(4)_R'$ symmetry is that even when the electroweak symmetry is broken spontaneously by the Higgs vev in \eq{ewsb}, there remain three exactly massless families of SM quarks.  This is easy to see if one redefines the $\Psi_{n2}$ fields at each of the black sites as $\Psi_{n2} = \Sigma^\dagger \Psi'_{n2}$; then $\CL_\text{sym}$ is independent of $\Sigma$, which means that the mass and Yukawa terms know nothing of electroweak symmetry breaking.  In effect, the SM families are forced to only have derivative couplings to the Higgs.   Therefore the three surplus RH singlet quarks cannot pair up with the three surplus LH doublet quarks, and one is left with three massless SM families. This mechanism differs from the flavor models  in which an approximate chiral flavor symmetry is responsible for keeping the SM families light ---  such as Minimal Flavor Violation models which start with a $U(3)^3$ symmetry among the quarks \cite{Chivukula:1987py}. To give the SM families mass requires breaking the $SU(4)$ symmetry, and to have mixing angles and nondegenerate quarks requires breaking the $U(3)$ symmetry; we do both with the same spurions at tree level.  However, the $SU(4)$ symmetry is also broken by radiative corrections in the form of $G_b$ gauge boson loops; this is an important issue but we defer discussion of that to \S~\ref{sec:radmass}.

\subsubsection{The $SU(4)\times U(3)$ symmetry breaking terms}

To give the SM quarks masses we introduce two spurions to break the $SU(4)\times U(3)$ symmetry, defined by the traceless $4\times 4$ matrices which can be thought of as transforming as elements of the adjoint of $SU(4)$:
\beq
X_u = \begin{pmatrix} 
1 & & &\cr
& 1 & &\cr
& & -3 &\cr
& & & 1
\end{pmatrix}\ ,\qquad
X_d = \begin{pmatrix} 
1 & & &\cr
& 1 & &\cr
& & 1 &\cr
& & & -3
\end{pmatrix}
\eeq
Both of these matrices break the $SU(4)$ symmetry down to $SU(3)\times U(1)$ and will allow the light fermions to acquire masses; the $X_u$ matrix splits off the $U$ quark from the $SU(4)$ multiplet, while $X_d$ distinguishes the $D$ quark.  We take for  our symmetry breaking mass terms
\beq
\CL_\text{asym} =\mybar \Psi_{m\alpha,L} \left[ \CM^{u}_{m\alpha,n\beta}+ \CM^{d}_{m\alpha,n\beta} \right]\Psi_{n\beta,R}  + h.c. 
\eqn{Lasym}\eeq
where
\beq
\CM^{u}_{m\alpha,n\beta} = M^u_{mn}
\otimes \begin{pmatrix}0 & 0\cr 0 & 1 \end{pmatrix}_{\alpha\beta}\otimes X_u \ ,\qquad
\CM^{d}_{m\alpha,n\beta} = M^d_{mn}
\otimes \begin{pmatrix}0 & 0\cr 0 & 1 \end{pmatrix}_{\alpha\beta}\otimes X_d \ .
\eeq
The $M^{u,d}$ matrices act on the $U(3)$ indices of the fermions, the structure of the $\{\alpha\beta\}$ matrix shows that only the Dirac fermions on black sites are involved, and  the $X$ matrices act on the implicit $SU(4)$ indices carried by each fermion.  By having the $X$ spurions each leave intact an $SU(3)$ subgroup of the black-site $SU(4)$ symmetry, we ensure that the fermions will not contribute any one-loop quadratically divergent mass contributions to the Higgs boson (the Little Higgs mechanism).  In fact, log divergences to the Higgs potential from one fermion loop also vanish in this model. 

The $M^{u,d}$ matrices in the above expression act on the indices of the three cells of our moose, explicitly breaking the $U(3)$ flavor symmetry, and we take them to have the textures
\beq
M^{u} = \begin{pmatrix} \CM^u_{11} & \CM^u_{12} & 0 \cr 0 & \CM^u_{22} &0 \cr \CM^u_{31} & 0 & \CM^u_{33} \end{pmatrix}\ \ \ ,\qquad
M^{d} = \begin{pmatrix} \CM^d_{11} &0& 0\cr \CM^d_{21} & \CM^d_{22} & 0 \cr 0&\CM^d_{32} &  \CM^d_{33} \end{pmatrix}\ .
\eeq
This choice has been made empirically, and we do not claim it to be unique, but these textures suggests the spurions could arise from a simple symmetry breaking scheme, which we do not pursue here. We will constrain all of the mass parameters to be real, except for $M^u_{31}$, whose phase will be the source of $CP$ violation in this model.    The diagonal elements break the $U(3)$ down to $U(1)^3$, allowing a nontrivial quark spectrum to emerge but no mixing angles; the off-diagonal terms will generate flavor mixing.

\section{A phenomenological fit}
\label{fit}

In order to study rare processes in a model which reproduces correctly the SM quark masses and mixing angles, we now fix 
\beq
M=5000\GeV\ ,\qquad f=1500\GeV\ ,\qquad \tan\beta = \frac{v_u}{v_d} = 1\ ,
\eeq
and fit the 11 real parameters plus one phase ($\lambda$,  and the $\CM^{u,d}$ matrices) to  the six quark masses, as well as the three mixing angles and one phase in the Cabibbo-Kobayashi-Maskawa matrix, a total of 10 data; our fit is neither unique, nor predictive in the SM quark sector, and the assumption of $\tan\beta=1$ is for simplicity, not following from any particular  Higgs potential.  In fact, one would expect $\tan\beta>1$ in these models, as discussed below, but considering different  values for $\tan\beta$ will not alter our analysis significantly. The point of this exercise is to produce a concrete model consistent with the SM in which we can accurately analyze low energy flavor phenomenology from new TeV physics. 

The fit we find  has
\beq
\lambda  =1.49794\ ,
\eeq
while the $\CM$ matrices (in GeV)   are given by
\beq
M^u=\left(
\begin{array}{ccc}
 1189.54 & 15.4904& 0 \\
 0 & 6.96490 & 0 \\
3.50799 e^{-i 1.224428} & 0 & 0.01441071
\end{array}
\right)\ ,\qquad
M^d=\left(
\begin{array}{ccc}
 45.7769 & 0 &0 \\
 -1.60269 & 0.600984 & 0 \\
 0 & 0.137582 & 0.0336607
\end{array}
\right)\ .
\eqn{cmfit}\eeq
These parameters allow us to reproduce the accepted values of the quark masses (in GeV), RG scaled to $\mu=1\TeV$ \cite{Xing:2007fb}:
\begin{equation}
\begin{aligned}
m_t&= 153.2& \quad
m_c &=5.32\times 10^{-1} &\quad
m_u &= 1.10\times 10^{-3} \cr
m_b&= 2.45& \quad
m_s &=4.69\times 10^{-2}& \quad
m_d &= 2.50\times 10^{-3} 
\end{aligned}
\end{equation}
and give rise to the CKM matrix
\beq
\left\vert V_\text{CKM}\right\vert =\left(
\begin{array}{ccc}
 0.974 & 0.226 & 0.00385 \\
 0.226 & 0.973 & 0.0423 \\
 0.00892 & 0.0415 & 0.998
\end{array}
\right)\eeq
and unitarity triangle angles
\beq
\sin(2\alpha) = 0.052\ ,\qquad
\sin(2\beta) = 0.72\ ,\qquad
\sin(2\gamma) = 0.68\ ,
\eeq
all values being within a few percent or better of the values given in Ref.~ \cite{PhysRevD.86.010001}.  

\begin{figure}[t]
\includegraphics[width=15cm]{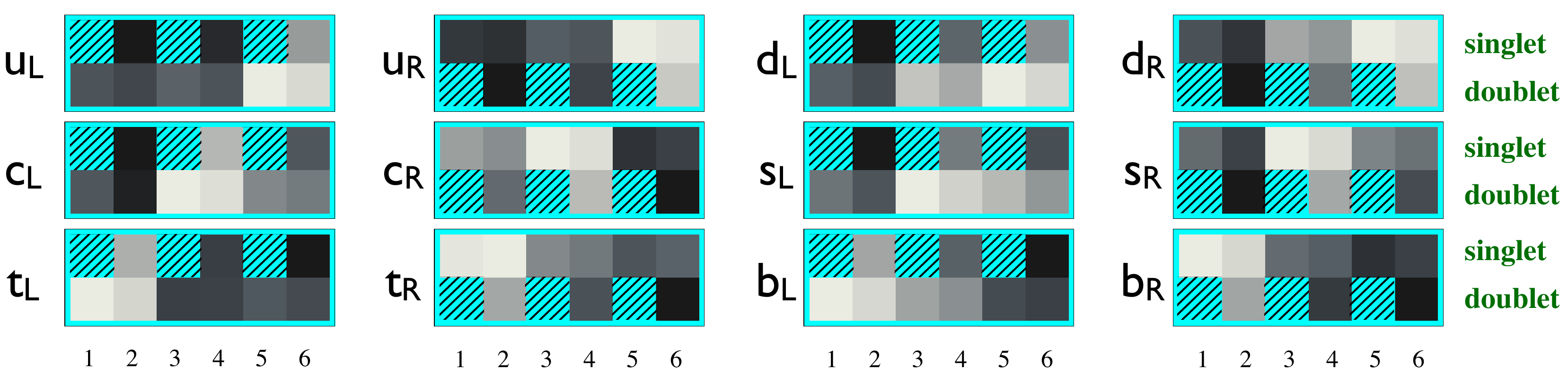}
\caption{ {A density plot of $\ln|\psi|^2$, where the  $\psi$  are the eigenvectors  of the LH and RH SM quark wavefunctions; the darker the square, the smaller the wavefunction.  Numbers $1,\ldots,6$ along the bottom indicate the site number in the moose of Fig.~\ref{fig:sixsite}; upper and lower rows indicate electroweak singlet and doublet components respectively.  One can see, for example, that families are mostly localized in different cells, with the LH down-type quarks being the most spread out, and that RH quarks have a little admixture of doublet, while LH quarks contain some singlet components. Hatched squares indicate combinations that do not exist in the model, such as a LH $SU(2)$-singlet up quark at site \#1.}}
\label{fig:wavefunctions}
\end{figure}

The wave functions for the SM quarks (i.e., their distribution over the six sites of the moose in Fig.~\ref{fig:sixsite}) can be visualized in Fig.~\ref{fig:wavefunctions}, where we provide a density plot of the $\ln|\psi|^2$. In this plot, light squares are where most of the support of the wavefunction is, and we see a clear pattern where each of the three families  resides mainly within its own cell of the moose.  This localization does not explain the mass hierarchy we achieve in this model:  that occurs because the $SU(4)$ symmetry in  \eq{cyval} allows the Higgs dependence to rotated out of the Yukawa couplings in $\CL_\text{sym}$, causing the Higgs to only couple through the $SU(4)\times U(3)$ violating spurion operators $\CM^{u,d}$ in $\CL_\text{asym}$, which have the hierarchy built into them (\eq{cmfit}).
However, the localization of families with small overlap in the extra dimension explains the smallness of FCNC in this model, since gauge boson couplings are local, and there are no large spurions breaking locality in this extra dimension which can be used to construct dangerous short-distance operators  from physics above the cutoff --- only the off-diagonal components of $\CM^{u,d}$ communicate between cells, and they are small. 

In addition to the SM quarks, the model contains six heavy exotic up and down quarks with masses given in (in $\TeV$)
\beq
U: &\quad& 6.628 ,\ 5.489 ,\ 5.482 ,\ 5.482 ,\ 5.463 ,\ 2.684 \cr
D: &\quad& 6.628 ,\  6.456 , \ 5.489 ,\  5.486 ,\ 5.482 , \ 5.482 \ .
\eeq
All of these masses are well below the cutoff of the effective theory, $\Lambda\sim 4\pi f \simeq 19\TeV$.

\subsection{Tree-level FCNC from the $Z$, $Z'$, and $Z''$ bosons}

We next consider the flavor properties of the neutral gauge bosons in the theory, the $Z$, $Z'$ and $Z''$.
All exotic gauge boson parameters depend on our choice for the angles $\gamma_{1,2}$, where $\gamma_i$ parametrizes the relative strength of the gauge interactions on the white and black sites respectively, as in \eq{gamma}, and   this section we make the somewhat arbitrary choice $\gamma_1 = \gamma_2=\pi/8$.  The $Z'$ and $Z''$ masses are then given by (\eq{heavyZ}) 
\beq
M_{Z'} = 750\GeV\ ,\qquad
M_{Z''} = 1400\GeV\qquad (\gamma_1=\gamma_2={\pi}/{8})\ .
\eqn{heavyZmasses}\eeq
Such masses would be ruled out by direct searches for new heavy neutral gauge bosons if the $Z'$ and $Z''$ had $Z$-like couplings to leptons; as we do not consider leptons in this paper, we simply assume that these two exotic gauge bosons are leptophobic; a more complete theory will have to address this issue.  Constraints on the flavor changing quark couplings of such bosons are relevant to this model, however, and we consider here the $\Delta F=1$ and $\Delta F=2$ processes arising from tree level neutral gauge boson exchange.

It is straightforward to compute the couplings of the gauge bosons for the phenomenological fit discussed above; it simply requires computing the currents coupling to the gauge boson mass eigenstates, and then substituting the light flavor eigenvectors for the $\Psi_{n\alpha a}$ fermions.  The results for the couplings of $Z, Z', Z'',W$ and  $W'$ are given in the appendix, \S~\ref{sec:GBcouplings}.

The off-diagonal neutral gauge boson couplings contribute to tree-level $\Delta S=2$ operators; in the case of the $Z$ we also have a tree-level contribution to the $\Delta S=1$ $K^0\to\mu^+\mu^-$ decay; however, from \eq{Z0coupling} we see that the   $\Delta S=1$ coupling to LH currents equals  $10^{-6}$ , which is sufficiently small to give a branching ratio several orders of magnitude below the observed branching ratio in this channel for the $K^0_L$. 

Squaring the largest $\Delta S=1$ couplings from \eq{Z0coupling}-\eq{Z2coupling} allows us to compute the coefficients of the  $\Delta S=2$ operators resulting from tree level $Z$, $Z'$ and $Z''$, with the results 
\beq
\frac{1\times 10^{-12}}{M_Z^2} \simeq \frac{1}{\left(10^5\TeV\right)^2}\ ,\qquad
\frac{4\times 10^{-10}}{M_{Z'}^2} \simeq \frac{1}{\left(4\times 10^4\TeV\right)^2}\ ,\qquad
\frac{1\times 10^{-8}}{M_{Z''}^2} \simeq \frac{1}{\left(1.3\times 10^4\TeV\right)^2}\ .
\eeq
The $Z$ and $Z'$ contributions are sufficiently small to have immeasurable  effects on kaon phenomenology; the $Z''$ contribution would be close to the current bounds if it were maximally $CP$ violating, but in fact the  phase in the $\mybar s d$ coupling of the $Z''$  is found to be $0.06$ in a basis where $V_{us}$ is real, so that its $\Delta S=2$ contributions are likewise compatible with experiment.  The product of left currents time right currents receives a chiral enhancement relative to left-left or right-right, but we find that the product of these couplings is very small in each case and not relevant.

\subsection{FCNC from physics above the cutoff}
As our theory is an effective theory for physics below the cutoff $\Lambda\simeq 4\pi f \simeq 19\TeV$, we need to consider whether dangerous FCNC effects can arise from contact operators arising from physics above the cutoff.  The generic power counting for operators in the effective theory is \cite{Cohen:1997rt}: (i) start with an overall factor of $\Lambda^2 f^2$; (ii) include a factor of $1/(\Lambda f^2)$ for each fermion bilinear in the operator; (iii) include a factor of $1/\Lambda$ for each derivative or $\CM$ spurion; (iii) include a factor of $1/f$ for each gauge field $A$ and a factor of $g/4\pi$ for each gauge generator $T$; (iv) include an overall dimensionless coupling assumed to be $O(1)$.   We first consider the example of operators contributing to $b\to s\gamma$ which are not a threat but which are simpler to analyze, before considering more sensitive $\Delta S=2$ operators for which there are stringent constraints.

\subsubsection{Example: tree level contributions to $b\to s\gamma$}
We first consider the most symmetric contact operators which could contribute  to $b\to s\gamma$,
\beq
\frac{c_1}{\Lambda^2}\left[(1+\delta)  \mybar \Psi_{m\alpha,L} \CM_{m\alpha,n\beta} \left(g_{2b}\tilde W^{\mu\nu}\sigma_{\mu\nu}\right) \Psi_{n\beta,R} 
+(1-\delta) \mybar \Psi_{m\alpha,L} \CM_{m\alpha,n\beta} \left(g_{1b} Y \tilde B^{\mu\nu} \sigma_{\mu\nu}\right)\Psi_{n\beta,R}+ h.c.\right]\ .
\eeq
These operators require an insertion of the $\CM$ spurion from \eq{cmval} which gives mass to the vectorlike fermions on the black sites and breaks their chiral $SU(4)$ symmetry down to the diagonal subgroup, as well as insertions of the gauge boson charges which break the vector $SU(4)$ symmetry further down to the gauged $SU(2)\times U(1)$.  The coefficients of the two operators should be the same up to radiative corrections, so we expect $c_1=O(1)$ while $\delta$ terms must actually arise from radiative corrections and involve three powers of the gauge generators instead of one, and hence be $O(\alpha/4\pi)$ by the power counting rules.

We can match the above interaction at tree level to the low energy operators 
\beq
\frac{1}{\Lambda^2}\left[\beta_1\frac{e m_b}{16\pi^2}\mybar b_L \sigma_{\mu\nu} s_R F^{\mu\nu} +  \beta_2 \frac{e m_b}{16\pi^2}\mybar b_R \sigma_{\mu\nu} s_L F^{\mu\nu} + h.c.\right]
\eeq
by expressing the $\Psi$ fields and the gauge fields in terms of mass eigenstates, and keeping only the light degrees of freedom of interest.  Using the solutions from \S\ref{fit} we find
\beq
|\beta_1| &=& |c_1|\bigl\vert(0.0129328 -0.0331439 i)-(56.9843 -145.866 i) \delta \bigr\vert\cr
|\beta_2| &=& |c_1|\bigl\vert(0.0267052 -0.0638978 i)+(32.3686 -83.9222 i) \delta \bigr\vert\ ,
\eeq
where the phases are a result of our choice of basis.  It is apparent from the above expression that the radiative correction proportional to $\delta = O(\alpha/4\pi)$ is comparable to the ``leading" term.   In either case, both contributions will be far smaller than SM contributions, since $\Lambda\simeq 19\TeV$.

Similarly, we can also consider operators involving insertions of $\CM^{u,d}$ instead of $\CM$ in the above operator, or operators that involve $\Psi$ on both black and white sites, such as
\beq
&&\frac{\lambda f}{\Lambda^2} \mybar \Psi_{m\alpha,L}\left(g_{1w}YB^{\mu\nu}+g_{2w}W^{\mu\nu}\right)\Sigma \CY_{m\alpha,n\beta}\sigma_{\mu\nu}\Psi_{n\beta}+\ldots
\eeq
where the ellipses refers to related terms involving the gauge fields at the black sites, as well as $(\CY \Sigma)^\dagger $ insertions.  In every case, the $1/\Lambda^2$ suppression makes these operators uninteresting compared to SM contributions.

\subsubsection{Contact operators contributing to $\Delta S=2$}

Next we consider $\Delta S=2$ four fermion operators, which will involve sums of products of two bilinear $\Delta S=1$ operators.  Therefore we perform the matching of $\Delta S=1$ bilinears of the form 
\beq
\mybar \Psi_{m\alpha a}\, S_{m\alpha a,n\beta b}\,\Gamma\, \Psi_{n\beta b}\to c\, \mybar s\, \Gamma\, d\ ,
\eeq
where $S$ is any spurion in the theory carrying both site and $SU(4)$ indices which are contracted with the fermion indices, made dimensionless with the appropriate powers of $\Lambda$ so that $c$ is dimensionless. $\Gamma$ is a Dirac matrix, and  we do not specify whether the operator is color singlet or color octet.  The $\Delta S=2$ operators will then be proportional to the square of such bilinears, with coefficient $c^2/\Lambda^2$.  We give here a list of  such a matching calculation  of $c$ for a variety of the largest contributions:
\medskip

\paragraph{$\mybar s_L d_R$}

\begin{equation}
\begin{aligned}
&\frac{1}{\Lambda}\mybar \Psi_L\CM\Psi_R: &\quad& |c |= 2\times 10^{-11}\\
&\frac{1}{\Lambda}\mybar \Psi_L\CM^u\Psi_R: & & |c |= 8\times 10^{-11}\\
&\frac{1}{\Lambda}\mybar \Psi_L\CM^d\Psi_R: & & |c |= 6\times 10^{-12}\\
&\frac{1}{\Lambda}\mybar \Psi_L(\CY \Sigma - \CY^\dagger \Sigma^\dagger)\Psi_R: & & |c |= 5\times 10^{-11}\\
\end{aligned}
\end{equation}
\paragraph{$\mybar s_R d_L$}
\begin{equation}
\begin{aligned}
&\frac{1}{\Lambda}\mybar \Psi_R\CM^\dagger\Psi_L: & \quad&|c |= 5\times 10^{-10}\\
&\frac{1}{\Lambda}\mybar \Psi_R(\CM^u)^\dagger\Psi_L: & &|c |= 8\times 10^{-10}\\
&\frac{1}{\Lambda}\mybar \Psi_R(\CM^d)^\dagger\Psi_L: & &|c |= 6\times 10^{-12}\\
&\frac{1}{\Lambda}\mybar \Psi_R(\CY \Sigma - \CY^\dagger \Sigma^\dagger)\Psi_L: & & |c |= 3\times 10^{-10}\\
\end{aligned}
\end{equation}
\paragraph{$\mybar s_L \gamma^\mu d_L$}
\begin{equation}
\begin{aligned}
&\frac{1}{\Lambda^2}\mybar \Psi_L\CM \CM^\dagger\gamma^\mu\Psi_L: &\quad & |c |= 8\times 10^{-6}\\
&\frac{1}{\Lambda^2}\mybar \Psi_L\CM^u \CM^\dagger\gamma^\mu\Psi_L: &\quad & |c |= 5\times 10^{-6}\\
&\frac{1}{\Lambda^2}\mybar \Psi_L\CM^d \CM^\dagger\gamma^\mu\Psi_L: &\quad & |c |= 3\times 10^{-7}\\
&\frac{1}{\Lambda^2}\mybar \Psi_L\CM^u (\CM^u)^\dagger\gamma^\mu\Psi_L: &\quad & |c |= 2\times 10^{-7}\\
&\frac{1}{\Lambda^2}\mybar \Psi_L(\CY \Sigma - \CY^\dagger \Sigma^\dagger)\CM^u\gamma^\mu\Psi_L: & & |c |= 5\times 10^{-6}\\
\end{aligned}
\end{equation}
\paragraph{$\mybar s_R  \gamma^\mu d_R$}
\begin{equation}
\begin{aligned}
&\frac{1}{\Lambda^2}\mybar \Psi_R \CM^\dagger \CM \gamma^\mu\Psi_R :&\quad& |c |= 1\times 10^{-6}\\
&\frac{1}{\Lambda^2}\mybar \Psi_R (\CM^u)^\dagger \CM \gamma^\mu\Psi_R :&\quad& |c |= 2\times 10^{-7}\\
&\frac{1}{\Lambda^2}\mybar \Psi_R (\CM^d)^\dagger \CM \gamma^\mu\Psi_R :&\quad& |c |= 9\times 10^{-7}\\
&\frac{1}{\Lambda^2}\mybar \Psi_R (\CM^u)^\dagger \CM^u \gamma^\mu\Psi_R :&\quad& |c |= 2\times 10^{-9}\\
&\frac{1}{\Lambda^2}\mybar \Psi_R(\CY \Sigma - \CY^\dagger \Sigma^\dagger)\CM^u\gamma^\mu\Psi_R: & & |c |= 2\times 10^{-7}\\
\end{aligned}
\end{equation}
Given that the $\Lambda\simeq 19\TeV$ in our model, and that the four fermion $\Delta S=2$ operators have a coefficient of $c^2/\Lambda^2$ (neglecting RG running effects) we find that all $\Delta S=2$ effects from short distance physics have a coefficient of  $\sim (2 \times 10^6 \TeV)^{-2}$ or smaller, and pose no problem for phenomenology.  The smallness of these operators cannot be attributed to having each family well localized within its own cell on the moose, since there exists sufficient overlap for a realistic Cabibbo angle.

\section{Radiative $SU(4)$ breaking corrections}
\label{sec:radmass}

In section \S~\ref{sec:yukawa} we discussed the important role played by the nonlinearly realized $SU(4)_R'$ symmetry of $\CL_\text{sym}$ in \eq{Lsym}, which enforced that the standard model families could only have derivative couplings to the Higgs.  This allowed us to introduce $SU(4)$ breaking soft spurions $\CM^{u,d}$ to give the families mass and distinguish between $u$-type and $d$-type quarks, along with $U(3)$ symmetry breaking which allowed us to generate nontrivial hierarchies and mixing angles.  A potential problem with this mechanism in the present model is that the $G_b$ gauge interactions explicitly break the $SU(4)_R'$ symmetry as well, and therefore we will have radiative corrections which spoil the symmetry. In particular, we expect  at one loop an $SU(4)_R'$ breaking radiative corrections to the mass $\CM$ in \eq{cmval} of form
\beq
\delta \CM \simeq  M\times\left[ \frac{ \alpha_{2b}}{4\pi}   \begin{pmatrix} 3/4 &&&\\ & 3/4 &&\\ &&0&\\&&&0\end{pmatrix} +  \frac{\alpha_{1b}}{4\pi}  \begin{pmatrix} 1/36&&&\\ &1/36 &&\\ &&4/9&\\&&&1/9\end{pmatrix} \right]\ ,
\eeq
where we took $M=5\TeV$ in our fit. One finds that even before turning on the $\CM^{u,d}$ spurions in \eq{Lasym}, this $SU(4)$ violating shift in $\CM$ gives a common mass to the standard model quarks of about $25\MeV$ or higher, depending on the strength of the $G_b$ gauge couplings, as parametrized by the angles $\gamma_{1,2}$ in \eq{gamma}.  This mass scales as $\sim 1/M$ for larger values of $M$, but is not very sensitive to reductions in $M$.

These radiative corrections are very interesting despite being bad news for our phenomenological model.  It provides a concrete example how particle masses can be generated radiatively, a dream of theorists since the discovery of the muon with mass $m_\mu\sim m_e/\alpha$.  However, since $25\MeV$ is roughly ten times larger than the up quark mass, this correction invalidates our phenomenological model as it stands.  There are several ways to address the problem in the  model:
\begin{enumerate}
\item
We could extend the black site gauge symmetry to $G_b=SU(2)\times SU(2)\times U(1)$ with a discrete symmetry forcing the two $SU(2)$ gauge couplings to have the same value.  At one loop the radiative corrections to $\CL_\text{sym}$ would be then be $SU(4)$ symmetric.  The $\Sigma$ field would spontaneously break $G_w \times G_b$ down to $SU(2)\times U(1)$ as before, but now there would be an additional massive $W''$ gauge boson which would eat the $\pi$ Goldstone boson.  The spurions $\CM^{u,d}$ would then have to actually be vacuum expectation values of  fields also spontaneously breaking $G_b\to SU(2)\times U(1)$.  In this case we would expect the $SU(4)$ violation to be communicated to the SM families with an additional $M_{W''}^2/M^2$ suppression, making the radiative contribution to quark masses at the $\sim 1\MeV$ level or smaller. The extended the $G_b$ gauge symmetry would also impact how the Higgs potential was constructed, but would not be hard to work around.
\item
Because of the see-saw nature of SM quark masses in our model, mixing through heavy Dirac families, raising the mass $M$ reduces the effect of radiative corrections.  Therefore we could make the $\CM$ operator $U(3)$ violating (but still $SU(4)$ symmetric) with larger values corresponding to sites where the lighter families sit.  In this way, $SU(4)$-violating  radiative corrections contributing to the lighter family masses would be reduced.  Presumably with such a hierarchy put in by hand  in $\CL_\text{sym}$, the hierarchy in the $\CM^{u,d}$ spurions could be less pronounced, but we have not pursued this.  
\item
It might also be possible to devise related models where the analogue of the radiative $SU(4)$ violation occurred only at two loops, which would render the effect negligibly small.
\end{enumerate}
We do not pursue these ideas further here, since the radiative correction problem does change the two most interesting features of this model: that a symmetry which is not a chiral family symmetry ($SU(4)_R'$ here) can enforce light SM family masses, and (ii) that it  is possible to have a phenomenologically sensible model with flavor at the $\TeV$ scale which does not invoke minimal flavor violation, and yet still does not have unacceptable FCNC.

\section{The $\Sigma$ field potential and vacuum alignment}

 The   pseudo-Nambu-Goldstone bosons (pNGBs) in this model parametrize the alignment of the vacuum, which determines whether or not the weak gauge bosons obtain mass via the Higgs mechanism. Some of the pNGBs are ``little'', meaning that their masses do not receive quadratically divergent one loop contributions from order one interactions,  and are naturally light  compared with the scale $f$. The little doublets, $H_u$ and $H_d$, serve as our little Higgs fields. In a successful little Higgs model these little Higgs doublets obtain a vev $v$ which is parametrically small compared with the compositeness scale $f$. 
Obtaining such  symmetry breaking pattern with a natural separation between $f$ and $v$ requires competing terms: larger  terms, which are minimized in the $\Sigma=1$ vacuum, but which begin at quartic powers of the little Higgs fields, and  smaller  terms, which begin at quadratic order in the little Higgs fields,  whose net effect is to slightly misalign the vacuum away from  $\Sigma=1$. In general the Yukawa  interactions  radiatively produce one loop finite, negative  quadratic terms and the gauge interactions produce smaller positive quadratic terms with a one loop log divergence.

\subsection{Radiative corrections and quadratic terms}

The divergence from the gauge loops may be absorbed into the counterterm for the effective interactions
\beq
V_{\rm eff}\supset c_{\rm gauge} \frac{f^4}{16\pi^2} \big[&&g_{2,w} ^2g_{2,b} ^2\sum_{a,c=1}^3\Tr\left( T_a   \Sigma T_c   \Sigma^\dagger \right)  \Tr\left(T_a\Sigma T_c \Sigma^\dagger \right)\cr
+&&g_{2,b}^2g_{1,w} ^2\sum_{a=1}^3\left(\Tr\left(  T_a \Sigma Y \Sigma^\dagger \right)\right)^2  \cr
+&&g_{1,b} ^2g_{2,w}^2 \sum_{a=1}^3\left(\Tr\left(Y\Sigma T_a\Sigma^\dagger \right)\right) ^2\cr+&& g_{1,b}^2 g_{1,w}^2\left(\Tr\left(Y \Sigma Y \Sigma^\dagger \right)\right)^2\big]
 \eqn{gaugepot}
 \eeq 
where the coefficient $c_{\rm gauge} $ is of order one and requires knowledge of the underlying theory to compute, but is assumed to be positive. The  interactions in \eq{gaugepot} give  mass of order $g^2 f/(4 \pi)$ to the $\pi^\pm$  and little Higgs fields. The field $H_U$ gets a large  finite negative contribution from the loops involving the top quark and its partners and small contributions from the other quarks, and $H_D$ gets small negative contributions.  In  a simple model with two top partners, an electroweak doublet with mass $m_L$ and an electroweak singlet with mass $m_R$, the one loop contribution to the Higgs mass squared would be:
\beq
\eqn{quarkpot}
\delta V_{\rm eff} \supset -\frac{3\lambda_t^2 H_U^\dagger H_U}{8 \pi^2} \frac{m_L^2m_R^2}{m_L^2-m_R^2}\log\frac{m_L^2}{m_R^2}\ .
\eeq
 In our model  there are a total of six partner quarks cooperating to cancel quadratic divergences from the top loop, but we checked numerically that  \eq{quarkpot}  holds to within 10\% when $m_L$ and $m_R$ are replaced with the masses of the two lightest exotic charge 2/3 quarks.  The negative quadratic terms for the little Higgs arise from a combination of the one-loop terms and additional small symmetry breaking terms which are introduced to give masses to all the pseudo-Nambu Goldstone bosons with parameters which may be tuned to give the electroweak scale in agreement with experiment. If we assume that the  126 GeV Higgs-like boson is the lightest boson in the Higgs sector and that it is standard model-like, then with our quark mass spectrum the various contributions to the quadratic term in Higgs potential cancel to within about 7\%, a mild tuning. 

\subsection{Plaquette terms}

In Ref. \cite{ArkaniHamed:2002qx} four sigma fields were introduced and ``plaquette" terms giving Higgs quartic interactions arose from combinations of terms involving traces of the four fields. This is obviously possible to repeat here, and there are no problems with the experimental viability of such a model. Since only one of the four sigma fields needs to couple to the fermions, introducing more fields will not affect the FCNC analysis. However, we wish to retain a more economical scalar sector for simplicity.  With only a single $\Sigma$ field, contributions to the Higgs potential may be introduced   using symmetry breaking spurions.    Note that a subset of the terms in \eq{gaugepot}, namely
 \beq 
 V_{\rm eff}\supset c_{\rm quartic} f^4 \left(\sum_{a=1}^3\left(\Tr\left(  T_a \Sigma Y \Sigma^\dagger \right)\right)^2   
+ \sum_{a=1}^3\left(\Tr\left(Y\Sigma T_a\Sigma^\dagger \right) \right)^2\right)
\eqn{gaugepot2}
\eeq 
have the feature that they begin at quartic order in the little Higgs fields, although obtaining an $O(1)$ quartic coupling requires a coefficient which is larger than the one induced by gauge loops, by a loop factor.  Other   terms inducing quartic but not quadratic terms  in the little Higgs fields are
 \beq \label{gaugepot3}V_{\rm eff}\supset  c'_{\rm quartic}f^4 \left(\sum_{a=1}^3\left(\Tr\left(  T_a \Sigma X_d \Sigma^\dagger \right)\right)^2   
+ \sum_{a=1}^3\left(\Tr\left(X_d\Sigma T_a\Sigma^\dagger \right) \right)^2\right)\eeq and
\beq \label{gaugepot4}V_{\rm eff}\supset  c''_{\rm quartic}f^4 \left(\sum_{a=1}^3\left(\Tr\left(  T_a \Sigma X_u \Sigma^\dagger \right)\right)^2   
+ \sum_{a=1}^3\left(\Tr\left(X_u\Sigma T_a\Sigma^\dagger \right) \right)^2\right)\eeq 

These terms do an adequate job of giving  a quartic potential for the neutral Higgs bosons. Unfortunately, unlike in some little Higgs models \cite{ArkaniHamed:2002qy,Schmaltz:2008vd,Hook:2009kx}, we do not have an underlying reason based on the gauge symmetry  for the inclusion of these terms and not others, which, with similar sized coefficients,    could give a Higgs mass term of order $f$. We note however that the terms which give a Higgs mass do not preserve the same subset of the global symmetries as the ones we have included, so their omission is technically natural. Renormalizing the theory  will require the introduction of spurions that could give a Higgs mass squared term, however with coefficients which can naturally be assumed to be suppressed by   loop factors.

 \subsection{Spurions contributing to the other scalar masses and vacuum alignment}
 
The term 
\beq
\eqn{chargeplaquette}
\Tr (X_u\Sigma X_d\Sigma^\dagger)+ \Tr (X_d\Sigma X_u\Sigma^\dagger) \eeq 
will give a mass to the $\pi^\pm$ and a quartic interaction involving charged Higgses, but no  terms involving only neutral components of $H_u$ or $H_d$.  Each of these terms preserves different $SU(3)$ symmetries under which the Higgses transform nonlinearly, and since the divergent parts of one loop diagrams only depend on a single interaction, the interaction \eq{chargeplaquette} will not lead to  one loop quadratic divergences in the Higgs potential. The term \eq{chargeplaquette}   can be used to make the charged Higgs bosons relatively heavy without quadratic divergences. With an $\CO(1)$ coefficient, this term will give masses to the $\pi^\pm$ of order $f$.
To build other gauge invariant spurions contributing to scalar masses and vacuum alignment, consider the field combinations
\beq
z_a =\Tr\left[ \begin{pmatrix} 0&\cr&\sigma_a\end{pmatrix} \Sigma\right]  
\eeq
for $a=0,\ldots,3$, where $\sigma_0$ is the unit matrix. 

The quantities   \beq
\CP_1= (|z_1|^2 +|z_2|^2)\ ,\quad
\CP_2=(z_1^2 +z_2^2)\ ,\quad
\eeq
 are gauge invariant and and begin at quartic order in the Higgs fields, and give mass to the $\pi^\pm$, but do not contribute to any quartic only involving neutral Higgses.
Other gauge invariant symmetry breaking terms are
\beq
 \CP_3= (|z_0|^2 +|z_3|^2)\ ,\quad
\CP_4=(z_0^2 -z_3^2)\ ,\quad
 \eeq
which begin at quadratic order in the Higgs fields. The term $\Re\CP_4$  gives a mass to the $\eta$, since it also contributes to the Higgs  masses.  If we keep the coefficient of this term small enough to   avoid  fine-tuning the Higgs mass, the $\eta$ mass will   be of order the weak scale or lighter. The term $\Im\CP_4$ violates parity and leads to a nonzero $\eta$ vev; its inclusion is optional and we will omit it to avoid this complication.

\section{Little flavor from extra dimensions}

The model described above was motivated by earlier work on the origin of families from extra dimensions, \cite{Kaplan:2011vz}.  Logically there is no need to consider the connection with extra dimensions, but we discuss the relation here  on the chance that it could lead to further development of either theory.

\subsection{The TI/domain wall fermion flavor mechanism}

A topological insulator (TI) is a material which has massless fermion surface modes, whose existence is dictated by topological properties of the fermion dispersion relation in the bulk of the material; for references, see \cite{2008PhRvB..78s5424Q,Hasan:2010xy}. The mechanism behind topological insulators is the same as that discovered earlier in the domain wall construction for lattice field theories in 4d with chiral fermions \cite{Kaplan:1992bt}.  A fascinating feature of such theories  is that the number of generations of light surface modes in these lattice theories can change discontinuously (for an semi-infinite material with a single surface) as  the coupling constants  in the underlying Lagrangian  are changed continuously, as first shown in \cite{Jansen:1992tw,Golterman:1992ub}.  These changes occur at critical couplings for which the bulk spectrum becomes gapless, at which point a winding number associated with  the fermion propagator jumps discontinuously from one value to another.   In particular, it was shown in  \cite{Jansen:1992tw,Golterman:1992ub} that when the Euclidian fermion propagator $S(p)$ is suitably regulated, the number of massless surface modes is a topological invariant proportional to the integral
\beq
\epsilon_{abcde} \int \frac{d^5p}{(2\pi)^5)} \Tr \left[S^{-1}(p)\partial_a S(p)\cdots S^{-1}(p)\partial_e S(p)\right]
\eeq
where the partial derivatives are with respect to the  5-momentum $p$, and the critical couplings at which the number of zeromodes can change are those for which the bulk gap vanishes and $S(p)$ develops a pole.  The idea presented in \cite{Kaplan:2011vz} was that the three generations of SM fermions observed in 4d could be such multiple surface modes of a single 5d bulk fermion, where this number of surface modes is determined by the topology of the 5d fermion dispersion relation, and an example was given which gave rise to three chiral families on the boundary of a semi-infinite extra dimension. 

There are several difficulties in implementing a realistic theory using this idea in its simplest form.  One is that while this mechanism can explain why the standard model has three families, it does not directly provide an explanation for the observed hierarchical structure of Yukawa couplings to the Higgs.  Another is that flavor physics is an inherently UV phenomenon in such models, and 5d field theories are not well defined in the UV \footnote{It is a UV phenomenon because the number of families depends on  the momentum space topology of the fermion dispersion relation, requiring knowledge of  the propagator $S(p)$ at large $p$.}. Finally, while three chiral families can arise when the 5d spacetime is semi-infinite with only one 4d surface, such a geometry is not compatible with observed gauge and gravitational interactions, as both gravitons and gauge fields necessarily live in the 5d bulk in such theories, and the bulk gauge fields are not compatible with 4d phenomenology.  If  the extra dimension is compactified to solve this problem, then fermion zeromodes generically appear in vector-like representations and cannot give rise to the observed chiral gauge theory of the SM at low energy.

As mentioned in \cite{Kaplan:2011vz}, the problem of chirality can be solved through a conventional $Z_2$ orbifold projection which we discuss below; the problem of UV ambiguity may be avoided by using the technique of deconstruction \cite{ArkaniHamed:2001nc}.    Combining the two gives rise to a class of theories such as the model discussed in this paper.

\subsection{The $Z_2$ orbifold projection}
Deconstruction replaces the extra dimension with a lattice; by treating gravity (and possibly gauge interactions) as strictly four-dimensional, deconstruction  yields a 4d theory with multiple copies of fields associated with the sites and links of the lattice.  Five dimensional locality translates into nearest neighbor interactions on this lattice, but is not required for the 4d theory to make sense.

In order  to ensure a chiral fermion spectrum, we require the action to be invariant under a $Z_2$ symmetry  under which all fields $\phi$ transform as $\phi\to \hat z \phi$ where $\hat z^2=1$.  The orbifold projection then consists of replacing every field $\phi$ in the model by 
\beq
\phi\to \CP_z \phi\ ,\qquad P_z = \half \left(1-\hat z\right)\ .
\eeq
The fields in our model consist of 5d (Dirac) fermions $\psi$ and gauge fields which live on sites, as well as bosonic link fields $\Sigma$ which will contain, among other mesons, the Higgs.   The action of the $Z_2$ on fermions is  
\beq
\hat z \psi_i=\CR_{ij}\gamma_5\psi_j
\eqn{fermz2} 
\eeq
with $\CR=\CR^\dagger$ and $\CR^2=1$.  
Hopping terms in the deconstructed 5d theory appear as mass terms in the 4d interpretation,
\beq
\mybar\psi_i M_{ij} P_R \psi_j + h.c.\ ,\qquad P_R = \half(1+\gamma_5)\ ,
\eqn{psimass}
\eeq
where $i,j$ are summed over sites and $M$ can be an arbitrary finite matrix so long as it respects the $Z_2$ symmetry,
\beq
-\CR M \CR = M\ .
\eqn{MZ2}
\eeq  

\begin{figure}[t]
\includegraphics[width=6 cm]{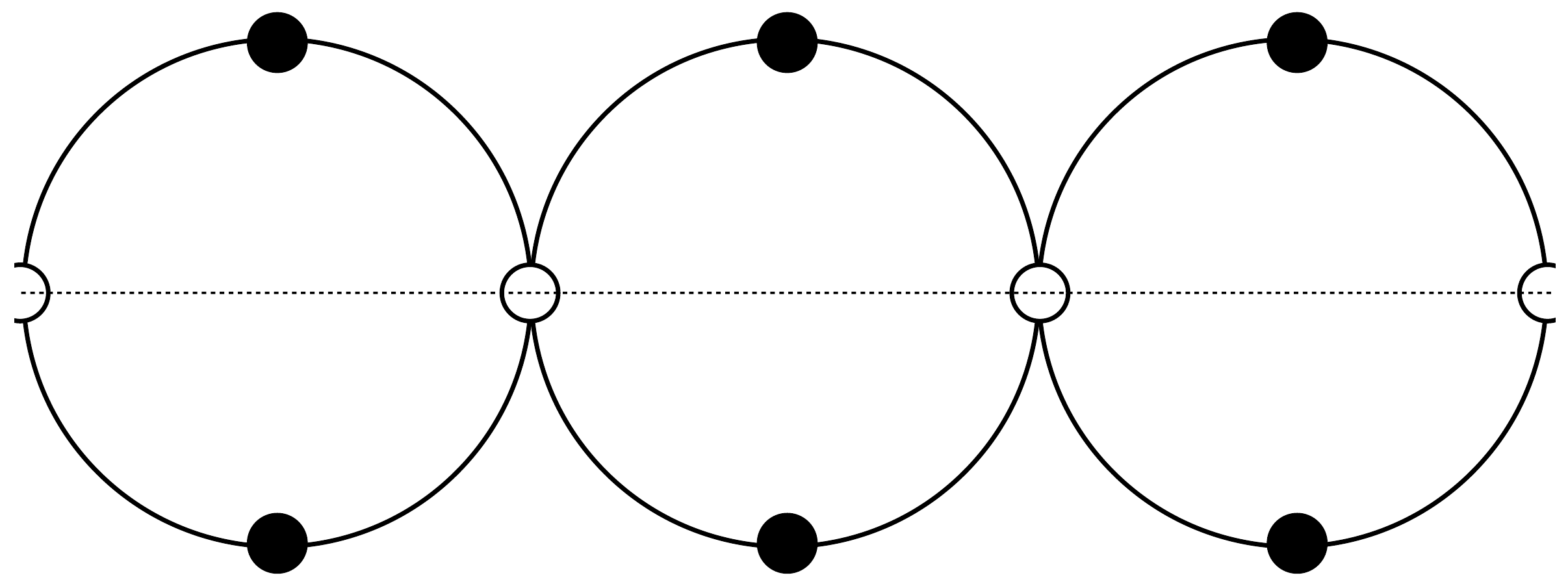}
\caption{{The lattice model prior to the $Z_2$ orbifold; the $Z_2$ symmetry acts as reflection about the horizontal axis and possesses three fixed points (white) and three pairs of points which transform into each other (black).  The arrangement is periodic, with the first and last white points identified.}}
\label{fig:orbi}
\end{figure}

An index theorem proved in the appendix (\S~\ref{sec:indexth}) states that
\beq
(\CN_L -\CN_R)= \Tr\CR\ ,
\eqn{index}\eeq
where $\CN_{L}$ and $\CN_R$ are the number of massless left-handed and right handed modes that survive the orbifold projection. If $R_{ij}$  represents a spatial reflection in the extra dimension taking site $i$ to site $j$, then nonzero diagonal elements in $R$ must equal $\pm1$ and are associated with the  fixed points of the $Z_2$ reflection.  The net number of chiral families thus equals the number of fixed points minus $2k$, where $k$ counts the number of $\{-1,1\}$ pairs of diagonal elements of $\CR$.  Since a simply connected curve will have an even number of fixed points under reflection, to obtain three standard model chiral families  will require exotic geometry in the extra dimension.  For example, we can consider the configuration and pictured in Fig.~\ref{fig:orbi} featuring an extra dimension in the shape of three circles arranged in a ring with three shared points (white dots in Fig.~\ref{fig:orbi}).  The action of $\CR$ is to reflect about the horizontal axis; black sites are exchanged while white sites are fixed points. 

We arrange the fermions on every site to be $4$s of $SU(4)$ 
\beq
\psi=\begin{pmatrix} u \\ d \\U\\D \end{pmatrix}
\eeq
where $u,d$ form an $SU(2)$ doublet and $U,D$ are $SU(2)$ singlets.  Then we specify that  the white site fermions are eigenstates of $\CR$, with $u,d$ having  eigenvalue $+1$ and $U,D$ having eigenvalue $-1$.  The orbifold projection therefore leaves  LH $SU(2)$ doublet zeromodes and RH $SU(2)$ singlet zeromodes at the white sites.  In contrast, $\CR$ interchanges the fermions at the pairs of black sites; the orbifold projection then reduces the two black sites to one, occupied by a single Dirac fermion.  The resulting theory looks like the moose of Fig~\ref{fig:sixsite} with fermion content of \eq{psi} and \eq{chi}.

The orbifold similarly reduces by half other fields that may live on the black sites, as well as link variables.
The way we included gauge fields and link variables  in the model discussed in this paper was motivated by a desire to keep the 4d model as simple as possible, and not to facilitate a 5d or higher dimensional  interpretation.  It is possible that other interesting models could be derived that are more faithful to a 5d spacetime interpretation, although the index theorem seems to require that the extra dimension be multiply connected if one requires three light families in the low energy theory.

\section{Discussion}

  In this model we have tied together the large global symmetries of little Higgs models with the global flavor symmetries that arise in a deconstruction of the extra-dimensional ``topological insulator" model of flavor in \cite{Kaplan:2011vz}. The role of these symmetries is different from any that have appeared previously in the flavor symmetry literature.  In particular, at tree level there is a nonlinearly realized $SU(4)$ symmetry which is not chiral and which ensures that the SM fermions only have derivative couplings to the Higgs at tree level. This symmetry is broken by radiative corrections which, along with the breaking of other symmetries at the few  $\TeV$ scale, allows us to generate realistic masses and mixing angles for the light fermions. Remarkably, even with the flavor physics at a few TeV,   flavor changing neutral currents from the new physics are smaller than those generated radiatively in the Standard Model.

The model described in this paper only describes the quarks. We expect that leptons may be included in a similar way, with care taken to ensure that light exotic vector mesons are leptophobic.  As with most flavor models involving leptons, a natural suppression mechanism for $\mu\to 3e$ will be critical. An explanation for the small size of the neutrino masses will require some new ingredient, such as a large Majorana mass for the right handed neutrinos.

We have not discussed the experimental signatures of the model, and, without having included the leptons, are not yet in a position to do so. An obvious signature is that the low energy effective theory below the TeV scale includes 2 Higgs doublets and a singlet, providing possible signatures in the usual searches for additional Higgs bosons.  Unlike in the original little Higgs models, which were very constrained by precision electroweak measurements, in this model the light quarks as well as the heavy quarks are linear combinations of quarks in transforming under different SU(2) and U(1) gauge groups, giving  partial cancellations in the coupling to the new gauge bosons. For example for the reference parameters considered here (such as $M=5\TeV$, $f=1.5\TeV$ and $\gamma_{1,2}=\pi/8$, which were not fine-tuned, plus the parameters $\lambda$ and $\CM_{u,d}$ chosen to correctly reproduce the quark masses and CKM angles, which were fine-tuned), the $Z'$ couplings to quarks are smaller  than the $Z$ couplings by a factor of around $10^{-2}$,  and the $Z''$ and $W'$ couplings are suppressed by about an order of magnitude relative to the $Z$ and $W$ couplings.  These suppressions are enough to satisfy current collider bounds for jet, top quark and gauge boson final states \cite{Chatrchyan:2012rva,CMS:2012xva,Aad:2012raa,Aad:2013wxa,Aad:2013nca,Chatrchyan:2013qha}. If we include the leptons in the obvious way, with Dirac neutrino masses, and no additional gauge groups, then the usual searches for new particles  decaying into leptons would constrain the model. In order to evade dilepton search constraints \cite{Chatrchyan:2012oaa,Aad:2012hf}, the new  neutral gauge bosons would have to be leptophobic,  even more weakly coupled to quarks,  or else  heavier. The couplings to quarks could be reduced further by tuning the mixing parameters $\gamma_i$.  Making the $W'$ and $Z',Z''$ bosons heavier without   increasing $f$ (which would increase the fine tuning) would require introducing another sigma field with a larger decay constant,   not coupled to fermions. The heavy gauge bosons would then eat the would be Nambu-Goldstone bosons from the other sigma field, leaving more  scalars light. Exploring these directions in model building and collider phenomenology is interesting but beyond the scope of this paper.

This model was constructed with the aim of providing a realistic detailed description of low energy phenomenology, so that a precise quantitive analysis of FCNC could be performed. As such, we focused on a numerical fit to low energy data, rather than taking a more qualitative and analytical approach.  Our fit does not predict the observed quark masses and mixing angles, as these are built into the structure of the spurions used, particularly $\CM^u$ and $\CM^d$ in \eq{cmfit}.  It  would be very interesting to be able to construct a more ambitious theory based on flavor symmetries which explained the structure of these spurions, and hence the SM particle spectrum, perhaps exploiting the radiative contribution to fermion masses discussed in \S~\ref{sec:radmass}.

 \begin{acknowledgments}

We would like to thank L. Dixon, H. Haber, M. Pospelov, and M. Schmaltz for useful conversations. D.K. and S.S. were supported in part by U.S.\ DOE grant No.\ DE-FG02-00ER41132;  S.S.  was  supported in part by the National Science Foundation under Grant No. PHY11-25915; A.N. was supported in part 
  by U.S.\ DOE grant No.\ DE-FG02-96ER40956.  S.S. gratefully acknowledges a graduate fellowship at the Kavli Institute for Theoretical Physics.
\end{acknowledgments}

\appendix

\section{Gauge boson couplings}
\label{sec:GBcouplings}
Here we give the gauge boson couplings to quarks, assuming the $[SU(2)\times U(1)]^2$ angles $\gamma_1=\gamma_2=\pi/8$ (see \eq{gamma}), using the fit parameters described in the text.  For these parameters, the masses of the gauge bosons are as given in \eq{heavyZmasses}, with $M_{Z'} = 750\GeV$ and $M_{Z''} = 1.4\TeV$.

\subsection{Neutral gauge boson couplings}

We parametrize the SM family parts of the neutral gauge boson currents in terms of four $3\times 3$ matrices for each vector meson as $\CL_V^{u,d}$ and $\CR_V^{u,d}$ where $V=\{Z, Z',Z''\}$ specifies the vector meson, $\CL$, $\CR$ indicates whether the current is LH or RH, and  $u,d$ specifies up-type versus down-type currents.  The results for our phenomenological fit
are as follows, where the basis is $\{u,c,t\}$ for the up-type quarks, and $\{d,s,b\}$ for the down-type quarks:
\begin{equation}
\begin{aligned}
&|\CL_Z^{u} |=\left(
\begin{array}{ccc}
 2.6\times 10^{-1} & 0 & 1.9\times 10^{-6} \\
 0 & 2.6\times 10^{-1} & 9.7\times 10^{-6} \\
 1.9\times 10^{-6} & 9.7\times 10^{-6} & 2.6\times 10^{-1}
\end{array}
\right)\ ,\qquad
&|\CR_Z^{u}| =\left(
\begin{array}{ccc}
 1.1\times 10^{-1} & 0 & 2.3\times 10^{-6} \\
 0 & 1.1\times 10^{-1} & 1.0\times 10^{-5} \\
 2.3\times 10^{-6} & 1.0\times 10^{-5} & 1.1\times 10^{-1}
\end{array}
\right)\ ,\\ &\\
&|\CL_Z^{d}| =\left(
\begin{array}{ccc}
 3.2\times 10^{-1} & 1.0\times 10^{-6} & 5.0\times 10^{-6} \\
 1.0\times 10^{-6} & 3.2\times 10^{-1} & 2.3\times 10^{-5} \\
 5.0\times 10^{-6} & 2.3\times 10^{-5} & 3.2\times 10^{-1}
\end{array}
\right)\ ,\qquad
&|\CR_Z^{d}| =\left(
\begin{array}{ccc}
 5.5\times 10^{-2} & 0 & 0 \\
 0 & 5.5\times 10^{-2} & 3.6\times 10^{-6} \\
 0 & 3.6\times 10^{-6} & 5.5\times 10^{-2}
\end{array}
\right) \ ,
\end{aligned}
\eqn{Z0coupling}\end{equation}

\begin{equation}
\begin{aligned}
& |\CL_{Z'}^{u}| = \left(
\begin{array}{ccc}
 2.6\times 10^{-3} & 0 & 0 \\
 0 & 2.6\times 10^{-3} & 3.4\times 10^{-5} \\
 0 & 3.4\times 10^{-5} & 3.8\times 10^{-3}
\end{array}
\right)\ ,\qquad
& |\CR_{Z'}^{u}| =\left(
\begin{array}{ccc}
 1.4\times 10^{-2} & 0 & 4.0\times 10^{-4} \\
 0 & 1.5\times 10^{-2} & 1.7\times 10^{-3} \\
 4.0\times 10^{-4} & 1.7\times 10^{-3} & 3.7\times 10^{-1}
\end{array}
\right) \ ,\\ &\\
& |\CL_{Z'}^{d}| =\left(
\begin{array}{ccc}
 5.\times 10^{-3} & 1.9\times 10^{-5} & 8.9\times 10^{-5} \\
 1.9\times 10^{-5} & 4.9\times 10^{-3} & 4.1\times 10^{-4} \\
 8.9\times 10^{-5} & 4.1\times 10^{-4} & 3.7\times 10^{-3}
\end{array}
\right) \ ,\qquad
& |\CR_{Z'}^{d}| =\left(
\begin{array}{ccc}
 6.7\times 10^{-3} & 0 & 2.6\times 10^{-5} \\
 0 & 6.6\times 10^{-3} & 2.0\times 10^{-4} \\
 2.6\times 10^{-5} & 2.0\times 10^{-4} & 8.8\times 10^{-3}
\end{array}
\right) \ ,
\end{aligned}
\eqn{Z1coupling}\end{equation}

\begin{equation}
\begin{aligned}
 &|\CL_{Z''}^{u} |= \left(
\begin{array}{ccc}
 1.9\times 10^{-2} & 0 & 7.9\times 10^{-5} \\
 0 & 1.9\times 10^{-2} & 2.8\times 10^{-4} \\
 7.9\times 10^{-5} & 2.8\times 10^{-4} & 2.9\times 10^{-2}
\end{array}
\right)\ ,\qquad
&|\CR_{Z''}^{u}| =\left(
\begin{array}{ccc}
 1.4\times 10^{-3} & 0 & 0 \\
 0 & 1.4\times 10^{-3} & 0 \\
 0 & 0 & 1.3\times 10^{-3}
\end{array}
\right) \ ,\\ &\\
&|\CL_{Z''}^{d} |= \left(
\begin{array}{ccc}
 2.0\times 10^{-2} & 1.0\times 10^{-4} & 5.0\times 10^{-4} \\
 1.0\times 10^{-4} & 1.9\times 10^{-2} & 2.3\times 10^{-3} \\
 5.0\times 10^{-4} & 2.3\times 10^{-3} & 2.9\times 10^{-2}
\end{array}
\right)\ ,\qquad
&|\CR_{Z''}^{d}| = \left(
\begin{array}{ccc}
 1.6\times 10^{-3} & 0 & 0 \\
 0 & 1.6\times 10^{-3} & 0 \\
 0 & 0 & 9.7\times 10^{-4}
\end{array}
\right)\ , \\
\end{aligned}
\eqn{Z2coupling}\end{equation}
For   legibility we have set to zero all entries smaller than $10^{-6}$, and only give the absolute values of the entries in the vector meson coupling matrices.
These couplings depend on the choice of $\gamma_{1,2}$; for $\gamma_{1,2} = \pi/5$ we find that the largest flavor diagonal coupling of the $Z'$ is bigger by about a factor of six, but remains smaller than the $Z$ diagonal couplings by about a factor of eight.

\subsection{Charged current couplings}

The $W$ boson mass has been fit to experiment, while the $W'$ boson is degenerate with the $Z''$, with a mass of $1.4\TeV$ for $\gamma_1=\gamma_2=\pi/8$.  We write the charged current couplings as
\beq
-\frac{g_2}{\sqrt{2}} \,\left({W}^+_\mu \bar u_i \left[\CL^{W}_{ij} P_L + \CR^{W}_{ij} P_R\right] d_j + {W'}^+_\mu \bar u_i \left[\CL^{W'}_{ij} P_L + \CR^{W'}_{ij} P_R\right] d_j\right)\, .
\eeq
For the parameters given in the text,  we find for the couplings
\begin{equation}
\begin{aligned}
& |\CL_{W} |= \left(
\begin{array}{ccc}
 9.7\times 10^{-1} & 2.3\times 10^{-1} & 3.8\times 10^{-3} \\
2.3\times 10^{-1} & 9.7\times 10^{-1}  & 4.2\times 10^{-2}\\
8.9\times 10^{-3} &4.1\times 10^{-2} &1.0
\end{array}
\right)\ ,\qquad
&|\CR_{W}| =\left(
\begin{array}{ccc}
 2.2\times 10^{-3} & 2.7\times 10^{-5} & 0 \\
2.7\times 10^{-5} &2.2\times 10^{-3}& 3.1\times 10^{-5} \\
 0 & 2.4\times 10^{-5} & 1.1\times 10^{-3}
\end{array}
\right) \ ,\\ &\\
& |\CL_{W'} |=\left(
\begin{array}{ccc}
5.7\times 10^{-2}& 1.3\times 10^{-2} & 1.4\times 10^{-5} \\
1.3\times 10^{-2} & 5.6\times 10^{-2} & 3.3\times 10^{-3} \\
 1.0\times 10^{-3} &4.4\times 10^{-3} & 8.9\times 10^{-2}
\end{array}
\right)\ ,\qquad
&|\CR_{W'}| =\left(
\begin{array}{ccc}
5.2\times 10^{-3} & 6.5\times 10^{-5} & 0 \\
6.4\times 10^{-5} & 5.3\times 10^{-3} &7.4\times 10^{-5} \\
 0 & 5.9\times 10^{-5} &2.9\times 10^{-3}
\end{array}
\right)\ , \\
\end{aligned}
\eqn{Wcoupling}\end{equation}
where we have set to zero all entries smaller than $10^{-5}$.
This normalization gives $\CL_W = V_\text{CKM}$. 

We see that the model predicts small $W$ couplings to right-handed currents; such couplings can lead to an $m_t/m_b$ enhancement  relative to the SM in the weak penguin graph contributing to $b\to s\gamma$,  but the above $|\CR_W|_{33}$  element is small enough to ensure that this enhancement does not cause conflict with experiment.
We see also see that the  $W'$ couplings to the SM fermions are quite small  and will not lead to problems with precision electroweak corrections.  Note that when every gauge field is rescaled by its coupling constant, the $W$ wavefunction is constant around the moose in Fig.~\ref{fig:sixsite}, while the $W'$ wavefunction alternates in sign between white and black sites; it is this sign alternation which causes strong cancellations in the coupling of the $W'$ to SM fermions.

\section{Index theorem}
\label{sec:indexth}

We prove here the assertion  in \eq{index}, which was also stated without proof in \cite{Kaplan:2011vz}. Define LH and RH eigenmodes
\beq
MM^\dagger \phi^i_L = \lambda_i\phi^i_L\ ,\qquad M^\dagger M \phi_R = \lambda_i\phi^i_R\ ,
\eqn{eigen}
\eeq
where $M$ is the $Z_2$ invariant mass of \eq{psimass} and \eq{MZ2}, and the eigenmodes are assumed to be normalized. The set of eigenvalues $\{\lambda_i\}$ are real and non-negative.  Then for $\lambda_i\ne 0$, we can choose $\phi_{L,R}^i$ to satisfy
\beq
\phi_R^i = \frac{1}{\sqrt{\lambda_i} }M^\dagger \phi_L^i\ ,\qquad
\phi_L^i = \frac{1}{\sqrt{\lambda_i} }M \phi_R^i\ .
\eqn{psirel}\eeq
From \eq{MZ2} it follows that $[\CR,M^\dagger M] = [\CR , M M^\dagger] = 0$; therefore we can choose $\CR$ to be diagonal in this same basis; furthermore, since $\CR^2=1$, its eigenvalues $r$ equal $\pm1$. 

The fermion fields $\psi_{L,R}$ are expanded in the eigenstates $\phi_{L,R}$ times LH or RH spinors.  Consider $\lambda_i\ne 0$ and $\hat z\psi_L^i = -r_i \psi_L^i$ with $r_i^2=1$ (recall that $\hat z = \CR\gamma_5$);  it follows from \eq{MZ2} and \eq{psirel} that $\hat z\psi_R^i = -r_i \psi_R^i$.  
Therefore if we define
\beq
z_L\equiv \Tr \hat z P_L\bigl\vert_{ \lambda\ne 0}\ ,\qquad
z_R\equiv \Tr \hat z P_R\bigl\vert_{ \lambda\ne 0}\ ,
\eeq
we know that
\beq
z_L=z_R\ .
\eqn{zequal}\eeq

Now consider the zeromode solutions to \eq{eigen}, with $\lambda_i=0$.  Let $\CN^\pm_{L,R}$ be the number of LH or RH zeromodes with $\hat z=\pm1$ respectively. Since $M$ is a finite matrix we have equal number of LH and RH zeromodes,  
\beq
\CN_L^+ + \CN_L^- = \CN_R^+ + \CN_R^- \equiv \CN 
\eqn{nequal}\eeq

 We also have
\beq
\Tr\CR &=&-\Tr \hat z P_L=  -\left(z_L +\CN_L^+-\CN_L^-\right)  \cr&&\cr
&=&+\Tr \hat z P_R
=+ \left(z_R +\CN_R^+-\CN_R^-\right)\ .
\eqn{creq}\eeq
Making use of  \eq{zequal},  \eq{nequal},  and \eq{creq}, we arrive at our index theorem \eq{index}, 
\beq
\left(\CN_L^- - \CN_R^-\right) = \Tr\CR\ .
\eeq
\bibliography{flavorbib}

\begin{thebibliography}{37}
\expandafter\ifx\csname natexlab\endcsname\relax\def\natexlab#1{#1}\fi
\expandafter\ifx\csname bibnamefont\endcsname\relax
  \def\bibnamefont#1{#1}\fi
\expandafter\ifx\csname bibfnamefont\endcsname\relax
  \def\bibfnamefont#1{#1}\fi
\expandafter\ifx\csname citenamefont\endcsname\relax
  \def\citenamefont#1{#1}\fi
\expandafter\ifx\csname url\endcsname\relax
  \def\url#1{\texttt{#1}}\fi
\expandafter\ifx\csname urlprefix\endcsname\relax\def\urlprefix{URL }\fi
\providecommand{\bibinfo}[2]{#2}
\providecommand{\eprint}[2][]{\url{#2}}

\bibitem[{\citenamefont{Froggatt and Nielsen}(1979)}]{Froggatt:1978nt}
\bibinfo{author}{\bibfnamefont{C.}~\bibnamefont{Froggatt}} \bibnamefont{and}
  \bibinfo{author}{\bibfnamefont{H.~B.} \bibnamefont{Nielsen}},
  \bibinfo{journal}{Nucl.Phys.} \textbf{\bibinfo{volume}{B147}},
  \bibinfo{pages}{277} (\bibinfo{year}{1979}).

\bibitem[{\citenamefont{Cohen et~al.}(1996)\citenamefont{Cohen, Kaplan, and
  Nelson}}]{Cohen:1996vb}
\bibinfo{author}{\bibfnamefont{A.~G.} \bibnamefont{Cohen}},
  \bibinfo{author}{\bibfnamefont{D.}~\bibnamefont{Kaplan}}, \bibnamefont{and}
  \bibinfo{author}{\bibfnamefont{A.}~\bibnamefont{Nelson}},
  \bibinfo{journal}{Phys.Lett.} \textbf{\bibinfo{volume}{B388}},
  \bibinfo{pages}{588} (\bibinfo{year}{1996}), \eprint{hep-ph/9607394}.

\bibitem[{\citenamefont{Arkani-Hamed
  et~al.}(2001{\natexlab{a}})\citenamefont{Arkani-Hamed, Cohen, and
  Georgi}}]{ArkaniHamed:2001nc}
\bibinfo{author}{\bibfnamefont{N.}~\bibnamefont{Arkani-Hamed}},
  \bibinfo{author}{\bibfnamefont{A.~G.} \bibnamefont{Cohen}}, \bibnamefont{and}
  \bibinfo{author}{\bibfnamefont{H.}~\bibnamefont{Georgi}},
  \bibinfo{journal}{Phys.Lett.} \textbf{\bibinfo{volume}{B513}},
  \bibinfo{pages}{232} (\bibinfo{year}{2001}{\natexlab{a}}),
  \eprint{hep-ph/0105239}.

\bibitem[{\citenamefont{Arkani-Hamed
  et~al.}(2002{\natexlab{a}})\citenamefont{Arkani-Hamed, Cohen, Katz, and
  Nelson}}]{ArkaniHamed:2002qy}
\bibinfo{author}{\bibfnamefont{N.}~\bibnamefont{Arkani-Hamed}},
  \bibinfo{author}{\bibfnamefont{A.}~\bibnamefont{Cohen}},
  \bibinfo{author}{\bibfnamefont{E.}~\bibnamefont{Katz}}, \bibnamefont{and}
  \bibinfo{author}{\bibfnamefont{A.}~\bibnamefont{Nelson}},
  \bibinfo{journal}{JHEP} \textbf{\bibinfo{volume}{0207}}, \bibinfo{pages}{034}
  (\bibinfo{year}{2002}{\natexlab{a}}), \eprint{hep-ph/0206021}.

\bibitem[{\citenamefont{Arkani-Hamed
  et~al.}(2002{\natexlab{b}})\citenamefont{Arkani-Hamed, Cohen, Katz, Nelson,
  Gregoire et~al.}}]{ArkaniHamed:2002qx}
\bibinfo{author}{\bibfnamefont{N.}~\bibnamefont{Arkani-Hamed}},
  \bibinfo{author}{\bibfnamefont{A.}~\bibnamefont{Cohen}},
  \bibinfo{author}{\bibfnamefont{E.}~\bibnamefont{Katz}},
  \bibinfo{author}{\bibfnamefont{A.}~\bibnamefont{Nelson}},
  \bibinfo{author}{\bibfnamefont{T.}~\bibnamefont{Gregoire}},
  \bibnamefont{et~al.}, \bibinfo{journal}{JHEP}
  \textbf{\bibinfo{volume}{0208}}, \bibinfo{pages}{021}
  (\bibinfo{year}{2002}{\natexlab{b}}), \eprint{hep-ph/0206020}.

\bibitem[{\citenamefont{Schmaltz and Tucker-Smith}(2005)}]{Schmaltz:2005ky}
\bibinfo{author}{\bibfnamefont{M.}~\bibnamefont{Schmaltz}} \bibnamefont{and}
  \bibinfo{author}{\bibfnamefont{D.}~\bibnamefont{Tucker-Smith}},
  \bibinfo{journal}{Ann.Rev.Nucl.Part.Sci.} \textbf{\bibinfo{volume}{55}},
  \bibinfo{pages}{229} (\bibinfo{year}{2005}), \eprint{hep-ph/0502182}.

\bibitem[{\citenamefont{Kaplan and Georgi}(1984)}]{Kaplan:1983fs}
\bibinfo{author}{\bibfnamefont{D.~B.} \bibnamefont{Kaplan}} \bibnamefont{and}
  \bibinfo{author}{\bibfnamefont{H.}~\bibnamefont{Georgi}},
  \bibinfo{journal}{Phys.Lett.} \textbf{\bibinfo{volume}{B136}},
  \bibinfo{pages}{183} (\bibinfo{year}{1984}).

\bibitem[{\citenamefont{Kaplan et~al.}(1984)\citenamefont{Kaplan, Georgi, and
  Dimopoulos}}]{Kaplan:1983sm}
\bibinfo{author}{\bibfnamefont{D.~B.} \bibnamefont{Kaplan}},
  \bibinfo{author}{\bibfnamefont{H.}~\bibnamefont{Georgi}}, \bibnamefont{and}
  \bibinfo{author}{\bibfnamefont{S.}~\bibnamefont{Dimopoulos}},
  \bibinfo{journal}{Phys.Lett.} \textbf{\bibinfo{volume}{B136}},
  \bibinfo{pages}{187} (\bibinfo{year}{1984}).

\bibitem[{\citenamefont{Georgi et~al.}(1984)\citenamefont{Georgi, Kaplan, and
  Galison}}]{Georgi:1984ef}
\bibinfo{author}{\bibfnamefont{H.}~\bibnamefont{Georgi}},
  \bibinfo{author}{\bibfnamefont{D.~B.} \bibnamefont{Kaplan}},
  \bibnamefont{and} \bibinfo{author}{\bibfnamefont{P.}~\bibnamefont{Galison}},
  \bibinfo{journal}{Phys.Lett.} \textbf{\bibinfo{volume}{B143}},
  \bibinfo{pages}{152} (\bibinfo{year}{1984}).

\bibitem[{\citenamefont{Georgi and Kaplan}(1984)}]{Georgi:1984af}
\bibinfo{author}{\bibfnamefont{H.}~\bibnamefont{Georgi}} \bibnamefont{and}
  \bibinfo{author}{\bibfnamefont{D.~B.} \bibnamefont{Kaplan}},
  \bibinfo{journal}{Phys.Lett.} \textbf{\bibinfo{volume}{B145}},
  \bibinfo{pages}{216} (\bibinfo{year}{1984}).

\bibitem[{\citenamefont{Dugan et~al.}(1985)\citenamefont{Dugan, Georgi, and
  Kaplan}}]{Dugan:1984hq}
\bibinfo{author}{\bibfnamefont{M.~J.} \bibnamefont{Dugan}},
  \bibinfo{author}{\bibfnamefont{H.}~\bibnamefont{Georgi}}, \bibnamefont{and}
  \bibinfo{author}{\bibfnamefont{D.~B.} \bibnamefont{Kaplan}},
  \bibinfo{journal}{Nucl.Phys.} \textbf{\bibinfo{volume}{B254}},
  \bibinfo{pages}{299} (\bibinfo{year}{1985}).

\bibitem[{\citenamefont{Chivukula and Georgi}(1987)}]{Chivukula:1987py}
\bibinfo{author}{\bibfnamefont{R.~S.} \bibnamefont{Chivukula}}
  \bibnamefont{and} \bibinfo{author}{\bibfnamefont{H.}~\bibnamefont{Georgi}},
  \bibinfo{journal}{Phys.Lett.} \textbf{\bibinfo{volume}{B188}},
  \bibinfo{pages}{99} (\bibinfo{year}{1987}).

\bibitem[{\citenamefont{Kaplan and Sun}(2012)}]{Kaplan:2011vz}
\bibinfo{author}{\bibfnamefont{D.~B.} \bibnamefont{Kaplan}} \bibnamefont{and}
  \bibinfo{author}{\bibfnamefont{S.}~\bibnamefont{Sun}},
  \bibinfo{journal}{Phys.Rev.Lett.} \textbf{\bibinfo{volume}{108}},
  \bibinfo{pages}{181807} (\bibinfo{year}{2012}), \eprint{1112.0302}.

\bibitem[{\citenamefont{Agashe and Contino}(2006)}]{Agashe:2005dk}
\bibinfo{author}{\bibfnamefont{K.}~\bibnamefont{Agashe}} \bibnamefont{and}
  \bibinfo{author}{\bibfnamefont{R.}~\bibnamefont{Contino}},
  \bibinfo{journal}{Nucl.Phys.} \textbf{\bibinfo{volume}{B742}},
  \bibinfo{pages}{59} (\bibinfo{year}{2006}), \eprint{hep-ph/0510164}.

\bibitem[{\citenamefont{Panico et~al.}(2012)\citenamefont{Panico, Redi, Tesi,
  and Wulzer}}]{Panico:2012uw}
\bibinfo{author}{\bibfnamefont{G.}~\bibnamefont{Panico}},
  \bibinfo{author}{\bibfnamefont{M.}~\bibnamefont{Redi}},
  \bibinfo{author}{\bibfnamefont{A.}~\bibnamefont{Tesi}}, \bibnamefont{and}
  \bibinfo{author}{\bibfnamefont{A.}~\bibnamefont{Wulzer}}
  (\bibinfo{year}{2012}), \eprint{1210.7114}.

\bibitem[{\citenamefont{Barbieri et~al.}(2012)\citenamefont{Barbieri, Buttazzo,
  Sala, Straub, and Tesi}}]{Barbieri:2012tu}
\bibinfo{author}{\bibfnamefont{R.}~\bibnamefont{Barbieri}},
  \bibinfo{author}{\bibfnamefont{D.}~\bibnamefont{Buttazzo}},
  \bibinfo{author}{\bibfnamefont{F.}~\bibnamefont{Sala}},
  \bibinfo{author}{\bibfnamefont{D.~M.} \bibnamefont{Straub}},
  \bibnamefont{and} \bibinfo{author}{\bibfnamefont{A.}~\bibnamefont{Tesi}}
  (\bibinfo{year}{2012}), \eprint{1211.5085}.

\bibitem[{\citenamefont{Xing et~al.}(2008)\citenamefont{Xing, Zhang, and
  Zhou}}]{Xing:2007fb}
\bibinfo{author}{\bibfnamefont{Z.-z.} \bibnamefont{Xing}},
  \bibinfo{author}{\bibfnamefont{H.}~\bibnamefont{Zhang}}, \bibnamefont{and}
  \bibinfo{author}{\bibfnamefont{S.}~\bibnamefont{Zhou}},
  \bibinfo{journal}{Phys.Rev.} \textbf{\bibinfo{volume}{D77}},
  \bibinfo{pages}{113016} (\bibinfo{year}{2008}), \eprint{0712.1419}.

\bibitem[{\citenamefont{Beringer et~al.}(2012)\citenamefont{Beringer, Arguin,
  Barnett, Copic, Dahl, Groom, Lin, Lys, Murayama, Wohl
  et~al.}}]{PhysRevD.86.010001}
\bibinfo{author}{\bibfnamefont{J.}~\bibnamefont{Beringer}},
  \bibinfo{author}{\bibfnamefont{J.~F.} \bibnamefont{Arguin}},
  \bibinfo{author}{\bibfnamefont{R.~M.} \bibnamefont{Barnett}},
  \bibinfo{author}{\bibfnamefont{K.}~\bibnamefont{Copic}},
  \bibinfo{author}{\bibfnamefont{O.}~\bibnamefont{Dahl}},
  \bibinfo{author}{\bibfnamefont{D.~E.} \bibnamefont{Groom}},
  \bibinfo{author}{\bibfnamefont{C.~J.} \bibnamefont{Lin}},
  \bibinfo{author}{\bibfnamefont{J.}~\bibnamefont{Lys}},
  \bibinfo{author}{\bibfnamefont{H.}~\bibnamefont{Murayama}},
  \bibinfo{author}{\bibfnamefont{C.~G.} \bibnamefont{Wohl}},
  \bibnamefont{et~al.} (\bibinfo{collaboration}{Particle Data Group}),
  \bibinfo{journal}{Phys. Rev. D} \textbf{\bibinfo{volume}{86}},
  \bibinfo{pages}{010001} (\bibinfo{year}{2012}),
  \urlprefix\url{http://link.aps.org/doi/10.1103/PhysRevD.86.010001}.

\bibitem[{\citenamefont{Cohen et~al.}(1997)\citenamefont{Cohen, Kaplan, and
  Nelson}}]{Cohen:1997rt}
\bibinfo{author}{\bibfnamefont{A.~G.} \bibnamefont{Cohen}},
  \bibinfo{author}{\bibfnamefont{D.~B.} \bibnamefont{Kaplan}},
  \bibnamefont{and} \bibinfo{author}{\bibfnamefont{A.~E.}
  \bibnamefont{Nelson}}, \bibinfo{journal}{Phys.Lett.}
  \textbf{\bibinfo{volume}{B412}}, \bibinfo{pages}{301} (\bibinfo{year}{1997}),
  \eprint{hep-ph/9706275}.

\bibitem[{\citenamefont{Schmaltz and Thaler}(2009)}]{Schmaltz:2008vd}
\bibinfo{author}{\bibfnamefont{M.}~\bibnamefont{Schmaltz}} \bibnamefont{and}
  \bibinfo{author}{\bibfnamefont{J.}~\bibnamefont{Thaler}},
  \bibinfo{journal}{JHEP} \textbf{\bibinfo{volume}{0903}}, \bibinfo{pages}{137}
  (\bibinfo{year}{2009}), \eprint{0812.2477}.

\bibitem[{\citenamefont{Hook and Wacker}(2010)}]{Hook:2009kx}
\bibinfo{author}{\bibfnamefont{A.}~\bibnamefont{Hook}} \bibnamefont{and}
  \bibinfo{author}{\bibfnamefont{J.~G.} \bibnamefont{Wacker}},
  \bibinfo{journal}{JHEP} \textbf{\bibinfo{volume}{1006}}, \bibinfo{pages}{041}
  (\bibinfo{year}{2010}), \eprint{0912.0937}.

\bibitem[{\citenamefont{{Qi} et~al.}(2008)\citenamefont{{Qi}, {Hughes}, and
  {Zhang}}}]{2008PhRvB..78s5424Q}
\bibinfo{author}{\bibfnamefont{X.-L.} \bibnamefont{{Qi}}},
  \bibinfo{author}{\bibfnamefont{T.~L.} \bibnamefont{{Hughes}}},
  \bibnamefont{and} \bibinfo{author}{\bibfnamefont{S.-C.}
  \bibnamefont{{Zhang}}}, \bibinfo{journal}{\prb}
  \textbf{\bibinfo{volume}{78}}, \bibinfo{eid}{195424} (\bibinfo{year}{2008}),
  \eprint{0802.3537}.

\bibitem[{\citenamefont{Hasan and Kane}(2010)}]{Hasan:2010xy}
\bibinfo{author}{\bibfnamefont{M.}~\bibnamefont{Hasan}} \bibnamefont{and}
  \bibinfo{author}{\bibfnamefont{C.}~\bibnamefont{Kane}},
  \bibinfo{journal}{Rev.Mod.Phys.} \textbf{\bibinfo{volume}{82}},
  \bibinfo{pages}{3045} (\bibinfo{year}{2010}), \eprint{1002.3895}.

\bibitem[{\citenamefont{Kaplan}(1992)}]{Kaplan:1992bt}
\bibinfo{author}{\bibfnamefont{D.~B.} \bibnamefont{Kaplan}},
  \bibinfo{journal}{Phys.Lett.} \textbf{\bibinfo{volume}{B288}},
  \bibinfo{pages}{342} (\bibinfo{year}{1992}), \eprint{hep-lat/9206013}.

\bibitem[{\citenamefont{Jansen and Schmaltz}(1992)}]{Jansen:1992tw}
\bibinfo{author}{\bibfnamefont{K.}~\bibnamefont{Jansen}} \bibnamefont{and}
  \bibinfo{author}{\bibfnamefont{M.}~\bibnamefont{Schmaltz}},
  \bibinfo{journal}{Phys.Lett.} \textbf{\bibinfo{volume}{B296}},
  \bibinfo{pages}{374} (\bibinfo{year}{1992}), \eprint{hep-lat/9209002}.

\bibitem[{\citenamefont{Golterman et~al.}(1993)\citenamefont{Golterman, Jansen,
  and Kaplan}}]{Golterman:1992ub}
\bibinfo{author}{\bibfnamefont{M.~F.} \bibnamefont{Golterman}},
  \bibinfo{author}{\bibfnamefont{K.}~\bibnamefont{Jansen}}, \bibnamefont{and}
  \bibinfo{author}{\bibfnamefont{D.~B.} \bibnamefont{Kaplan}},
  \bibinfo{journal}{Phys.Lett.} \textbf{\bibinfo{volume}{B301}},
  \bibinfo{pages}{219} (\bibinfo{year}{1993}), \eprint{hep-lat/9209003}.

\bibitem[{\citenamefont{Chatrchyan
  et~al.}(2013{\natexlab{a}})}]{Chatrchyan:2012rva}
\bibinfo{author}{\bibfnamefont{S.}~\bibnamefont{Chatrchyan}}
  \bibnamefont{et~al.} (\bibinfo{collaboration}{CMS Collaboration}),
  \bibinfo{journal}{JHEP} \textbf{\bibinfo{volume}{1302}}, \bibinfo{pages}{036}
  (\bibinfo{year}{2013}{\natexlab{a}}), \eprint{1211.5779}.

\bibitem[{\citenamefont{Collaboration}(2012)}]{CMS:2012xva}
\bibinfo{author}{\bibfnamefont{CMS}~\bibnamefont{Collaboration}},
  \bibinfo{journal}{CERN Report No. CMS-PAS-EXO-11-093}
  (\bibinfo{year}{2012}).

\bibitem[{\citenamefont{Aad et~al.}(2013{\natexlab{a}})}]{Aad:2012raa}
\bibinfo{author}{\bibfnamefont{G.}~\bibnamefont{Aad}} \bibnamefont{et~al.}
  (\bibinfo{collaboration}{ATLAS Collaboration}), \bibinfo{journal}{JHEP}
  \textbf{\bibinfo{volume}{1301}}, \bibinfo{pages}{116}
  (\bibinfo{year}{2013}{\natexlab{a}}), \eprint{1211.2202}.

\bibitem[{\citenamefont{Aad et~al.}(2013{\natexlab{b}})}]{Aad:2013wxa}
\bibinfo{author}{\bibfnamefont{G.}~\bibnamefont{Aad}} \bibnamefont{et~al.}
  (\bibinfo{collaboration}{ATLAS Collaboration})
  (\bibinfo{year}{2013}{\natexlab{b}}), \eprint{1305.0125}.

\bibitem[{\citenamefont{Aad et~al.}(2013{\natexlab{c}})}]{Aad:2013nca}
\bibinfo{author}{\bibfnamefont{G.}~\bibnamefont{Aad}} \bibnamefont{et~al.}
  (\bibinfo{collaboration}{ATLAS Collaboration})
  (\bibinfo{year}{2013}{\natexlab{c}}), \eprint{1305.2756}.

\bibitem[{\citenamefont{Chatrchyan
  et~al.}(2013{\natexlab{b}})}]{Chatrchyan:2013qha}
\bibinfo{author}{\bibfnamefont{S.}~\bibnamefont{Chatrchyan}}
  \bibnamefont{et~al.} (\bibinfo{collaboration}{CMS Collaboration})
  (\bibinfo{year}{2013}{\natexlab{b}}), \eprint{1302.4794}.

\bibitem[{\citenamefont{Chatrchyan
  et~al.}(2013{\natexlab{c}})}]{Chatrchyan:2012oaa}
\bibinfo{author}{\bibfnamefont{S.}~\bibnamefont{Chatrchyan}}
  \bibnamefont{et~al.} (\bibinfo{collaboration}{CMS Collaboration}),
  \bibinfo{journal}{Phys.Lett.} \textbf{\bibinfo{volume}{B720}},
  \bibinfo{pages}{63} (\bibinfo{year}{2013}{\natexlab{c}}), \eprint{1212.6175}.

\bibitem[{\citenamefont{Aad et~al.}(2012)}]{Aad:2012hf}
\bibinfo{author}{\bibfnamefont{G.}~\bibnamefont{Aad}} \bibnamefont{et~al.}
  (\bibinfo{collaboration}{ATLAS Collaboration}), \bibinfo{journal}{JHEP}
  \textbf{\bibinfo{volume}{1211}}, \bibinfo{pages}{138} (\bibinfo{year}{2012}),
  \eprint{1209.2535}.

\bibitem[{\citenamefont{Nelson}(1988)}]{Nelson:1988wn}
\bibinfo{author}{\bibfnamefont{A.~E.} \bibnamefont{Nelson}},
  \bibinfo{journal}{Phys.Rev.} \textbf{\bibinfo{volume}{D38}},
  \bibinfo{pages}{2875} (\bibinfo{year}{1988}).

\bibitem[{\citenamefont{Arkani-Hamed et~al.}(2000)\citenamefont{Arkani-Hamed,
  Hall, Tucker-Smith, and Weiner}}]{ArkaniHamed:1999yy}
\bibinfo{author}{\bibfnamefont{N.}~\bibnamefont{Arkani-Hamed}},
  \bibinfo{author}{\bibfnamefont{L.~J.} \bibnamefont{Hall}},
  \bibinfo{author}{\bibfnamefont{D.}~\bibnamefont{Tucker-Smith}},
  \bibnamefont{and} \bibinfo{author}{\bibfnamefont{N.}~\bibnamefont{Weiner}},
  \bibinfo{journal}{Phys.Rev.} \textbf{\bibinfo{volume}{D61}},
  \bibinfo{pages}{116003} (\bibinfo{year}{2000}), \eprint{hep-ph/9909326}.

\bibitem[{\citenamefont{Arkani-Hamed
  et~al.}(2001{\natexlab{b}})\citenamefont{Arkani-Hamed, Cohen, and
  Georgi}}]{ArkaniHamed:2001ca}
\bibinfo{author}{\bibfnamefont{N.}~\bibnamefont{Arkani-Hamed}},
  \bibinfo{author}{\bibfnamefont{A.~G.} \bibnamefont{Cohen}}, \bibnamefont{and}
  \bibinfo{author}{\bibfnamefont{H.}~\bibnamefont{Georgi}},
  \bibinfo{journal}{Phys.Rev.Lett.} \textbf{\bibinfo{volume}{86}},
  \bibinfo{pages}{4757} (\bibinfo{year}{2001}{\natexlab{b}}),
  \eprint{hep-th/0104005}.

\end{thebibliography}
\end{document}